\input amstex
\documentstyle{amsppt}
\magnification=1200
\font\cyr=wncyr10
\font\cyb=wncyb10
\font\cyi=wncyi10
\font\cyre=wncyr8
\font\cyie=wncyi8
\NoBlackBoxes
\NoRunningHeads
\define\id{\operatorname{\bold i\bold d}}
\define\res{\operatorname{res}}
\define\PSLTWO{\operatorname{\bold P\bold S\bold L}(2,\Bbb C)}
\define\sltwo{\operatorname{\frak s\frak l}(2,\Bbb C)}
\define\ad{\operatorname{ad}}
\define\Vrt{\operatorname{Vert}}
\define\Mat{\operatorname{Mat}}
\define\SU{\operatorname{\bold S\bold U}}
\define\Vect{\operatorname{Vect}}
\define\DOP{\operatorname{DOP}}
\define\HS{\operatorname{\Cal H\Cal S}}
\define\Map{\operatorname{Map}}
\define\Diff{\operatorname{Diff}}
\define\End{\operatorname{End}}
\define\Rot{\operatorname{Rot}}
\document\qquad\qquad\qquad\qquad\qquad\qquad\qquad\qquad\qquad\qquad
$\boxed{\boxed{\aligned
&\text{\eightpoint"Thalassa Aitheria" Reports}\\
&\text{\eightpoint RCMPI-96/05$^+$}\endaligned}}$\newline
\ \newline
\ \newline
\topmatter
\title Droems: experimental mathematics, informatics and infinite
dimensional geometry\endtitle
\author\rm D.V.Juriev\footnote{\ This is an English translation of the
original Russian version, which is located at the end of the article as an 
appendix. I regret that the translation is too far from the precision
of the original so in a case of any differences between English and Russian
versions caused by a translation the least has the priority as the original
one. A remark is convenient: the term ``informatics'' widely adopted in
Russian literature as for ``computer science'' as for ``information science''
and sometimes for the whole cybernetics (as it was proposed by N.Wiener)
is used in the article without any specification of its meaning. Some other
terms of Latin form common in Russian literature (e.g. of French origin) are
used instead of their sometimes slightly different English equivalents.
The same is true also for the terms of a Greek root. Moreover, some
purely Russian terms are directly transliterated in English if their
translation is not adopted and any variant is not adequate.\newline}
\endauthor
\affil\eightpoint{``Thalassa Aitheria'' Research Center for Mathematical
Physics and Informatics\linebreak\rm {\it E-mail:\/} denis\@juriev.msk.ru}
\endaffil
\endtopmatter
\document
This paper being addressed as to mathematicians-theorists specializing
in geometry, algebra, functional analysis, theory of dynamical and
controlled systems and interested in modern applications of their
disciplines to problems of information technologies (an organization
of real-time interactive dynamical videosystems for accelerated
computer and telecommunications) as to specialists, who elaborate 
such systems with an accent on mathematical methods of interactive computer
graphics in real time, is devoted to an analysis of a certain apparently
rather perspective as well as an interesting from the pure (as experimental
as theoretical) mathematics point of view approach to solution of one
problem of the information technologies.

The main difficulty that lays obstacles to the further development of
real-time interactive systems for accelerated computer and
telecommunications is the essential scantiness of a volume of information
transmitted per unit of time. If for the non-interactive systems and for
the conventional time interactive systems arranged for accelerated
communications this difficulty may be more or less overcame by
the using of various methods of compression and encoding as well as
by the simultaneous development of the matherial (technical) basis
then for the real-time interactive systems it becomes almost
insurmountable under a naive approach. Moreover, the using of
the real-time interactive systems for accelerated communications on
large distances produces a new difficulty related to the impossibility
to synchronize the internal rythms of subjects of a communication 
caused by the relativistic effects; thus, if the distance is about
$N$ thousand of kilometers then a deviation in synchronization is
theoretically not less than $N/300$ second and practically much more.

All these difficulties produce a necessity in a new organization
of information and its transmission in the real-time interactive
systems for accelerated communications. One of the possibilities,
by the way used in systems of ordinary communications, is in a
transmission the pointers to objects instead of their descriptors
with the following dynamical reconstruction of the objects themselves.
Under the using of this possibility in the interactive real-time
psychoinformation videosystems for accelerated nonverbal cognitive
communications, which seem to be the most perspective to the author
by many reasons, which will be clarified below, the connection
between pointers and objects themselves as well as the process of
dynamical reconstruction performed in real time and as a sequence
at least partially subconsciously that realizes its characteristic
feature, which differs the dynamical reconstruction of droems from
ordinary reconstruction-``decoding'', become nontrivial, potentially
very pithy but almost unexplored as experimentally as theoretically.
However, it is namely the knowledge of mathematical foundations,
which lie in their basis, may provide with reduction of excessive
claims to the apparate base of the systems allowing to realize them
by widely available sources and on another hand may be a
startpoint for elaboration of convenient and relatively simple
software.

In view of all these circumstances this article is an attempt
to explicate the mathematical foundations for the prescribed approach to
organization of information and its transmission in the interactive
real-time psychoinformation videosystems for accelerated nonverbal
cognitive communications; on this way the key role is played by
the droems (``dynamically reconstructed objects of experimental
mathematics'') and interpretational figures as pointers to them.

The article is organized in the following manner: four paragraphs
are devoted to (1) an exposition of basic notions of the interpretational
geometry, (2) the operator methods in the theory of interactive dynamical
videosystems, (3) the general concept of the organization of integrated
interactive real-time videocognitive systems, (4) the dynamically
reconstructed objects of experimental mathematics and processes
of their dynamical reconstruction, where the general notions are illustrated
by a concrete example related to the infinite dimensional geometry.
The exposition is presumably heuristic and conceptual (the first and
the third paragraphs) though some particular aspects such as
content of the second and the fourth paragraphs, which allow deeper
formalization and detailing in present, are exposed on the mathematical
level of rigor.

Note also that the motivations for particular statements exposed in
the article lie in the concrete solutions used in various learning
and communication systems, however, the consistent and systematic
experimentally mathematical approach to the theme is contained in
literature for the first time.

\head\bf\S 1. Interpretational geometry\endhead

This paragraph is devoted to an exposition of various aspects of
interpretational geometry, which have a relation to the organization
of information and its transmission in interactive videosystems of
computer and telecommunications. The main attention is paid to
interpretational figures, which will be of interest below as
pointers to droems, and to forms of their transmission, otherwords,
to the specific features of interpretational geometry in the
multi-user mode.

\subhead 1.1. Interpretational figures [1,2:App.A]\endsubhead
Geometry described below is related to a class of interactive
information systems. Let us call an interactive information system
computer graphic (or interactive information videosystem) if the
information stream ``computer--user'' is organized as a stream of
geometric graphical data on a screen of monitor; an interactive
information system will be called psychoinformation if an information
transmitted by the channel ``user--computer'' is (completely or
partially) subconscious. Thus, in the interactive systems that we are
interested in the control is coupled with unknown or incompletely
known feedback, systems with such control will be called
{\it interactively controlled}. In general, an investigation of
interactively controlled (psychoinformation) systems for an
experimental and a theoretical explication of possibilities
contained in them, which are interesting for mathematical sciences
themselves, and of ``hidden'' abstract mathematical objects, whose
observation and analysis are actually and potentially realizable
by these possibilities, is an important problem itself. So below
there will be defined the notions of an interpretational figure
and its symbolic drawing that undoubtly play a key role in the
description of a computer--geometric representation of mathematical
data in interactive information systems. Below, however, the accents
will be focused a bit more on applications to informatics preserving
a general experimentally mathematical view, the interpretational
figures (see below) will be used as pointers to droems and interactive
real-time psychoinformation videosystems will be regarded as components
of integrated interactive videocognitive systems for accelerated
nonverbal cognitive communications.

In interactive information systems mathematical data exist in the
form of an interrelation between the geometric internal image (figure)
in the subjective space of the observer and the computer-graphic external
representation. The latter includes visible (drawings of the figure)
and invisible (analytic expressions and algorithms for constructing
these images) elements. Identifying geometric images (figures) in
the internal space of the observer with computer-graphic representations
(visible and invisible elements) is called a {\it translation}, in this
way the visible object may be not identical with the figure, so that
separate visible elements may be considered as modules whose translation
is realized independently. The translation is called an {\it interpretation}
if the translation of separate modules is performed depending on the
results of the translation of preceding ones.

\definition{Definition 1} The figure obtained as a result of
interpretation is called an {\it interpretational figure}.
\enddefinition

Note that the interpretational figure may have no usual formal
definition; namely, only if the process of interpretation admits an
equivalent process of compilation definition of the figure is reduced
to definitions of its drawings that is not true in general. So the
drawing of an interpretational figure defines only dynamical ``technology
of visual perception'' but not its ``image'', such drawings will be
called {\it symbolic}.

The computer-geometric description of mathematical data in interactive
information systems is closely connected with the concept of anomalous
virtual reality.
\pagebreak

\subhead 1.2. Intentional anomalous virtual realities [1,2:App.A]\endsubhead

\definition{Definition 2 {\rm (cf.[1,2:App.A])}}
{\bf (A).} {\it Anomalous virtual reality\/} ({\it AVR\/}) {\it in a narrow
sense\/} means some system of rules of a nonstandard descriptive geometry
adapted for realization on videocomputers (or multisensorial systems of
``virtual reality'' [3-6]). {\it Anomalous virtual reality in a wide
sense\/} also involves an image in cyberspace formed in accordance with
said system of rules. We shall use the term in its narrow sense.
{\bf (B).} {\it Naturalization\/} is the constructing of an AVR from
some abstract geometry or physical model. We say that anomalous virtual
reality {\it naturalizes\/} the abstract model and the model {\it
transcendizes\/} the naturalizing anomalous virtual reality.
{\bf (C).} {\it Visualization\/} is the constructing of certain image
or visual dynamics in some anomalous virtual reality (realized by
hardware and software of a computer-grafic interface of the concrete
videosystem) from the objects of an abstract geometry or processes in
a physical model.
{\bf (D).} Anomalous virtual reality, whose objects depend on the observer,
is called an {\it intentional anommalous virtual reality\/} ({\it IAVR\/}).
the generalized perspective laws for IAVR contain the interactive dynamical
equations for the observed objects in addition to standard (geometric)
perspective laws. In IAVR the observation process consists of a physical
process of observation and a virtual process of intentional governing of
the evolution of images in accordance with the dynamical perspective laws.
\enddefinition

In intentional anomalous virtual reality (IAVR) that is realized
by hardware and software of the {\it computer-graphic interface of
the interactive videosystem} being geometrically modelled by this
IAVR (on the level of descriptive geometry whereas the model transcendizing
this IAVR realizes the same on the level of abstract geometry) respectively,
the observed objects are demonstrated as connected with the observer who
acts on them and determines, or fixes, their observed states so that
the objects are thought only as a potentiality of states from the given
spectrum whose realization depends also on the observer. The symbolic
drawings of interpretational figures may be considered as states of
some IAVR.

Note that mathematical theory of anomalous virtual realities (AVR)
including the basic procedures of naturalization and thanscending
connected AVR with the abstract geometry is a specific branch of {\it
modern nonclassical descriptive (computer) geometry}.

\definition{Definition 2D} The set of all continuously distributed
visual charcteristics of the image in anomalous virtual reality is called
an {\it anomalous color space\/}; the anomalous color space elements
of noncolor nature are called {\it overcolors}, and the quantities
transcendizing them in an abstract model are called {\it ``latent
lights''}. The set of the generalized perspective laws in a fixed
anomalous color space is called a {\it color-perspective system\/};
two AVRs are called equivalent if their color-perspective systems
coincide; AVR equivalent to one realized on the videocomputer (but not
realized itself) is called {\it marginal}.
\enddefinition

\subhead\bf 1.3. Non-Alexandrian interpretational geometry [7,2:App.A]
\endsubhead Note that the majority of classical geometries use a postulate
that we shall call an Alexandrian postulate but do not include it
explicitely in their axiomatics.

\definition{\bf The Alexandrian postulate} \it Any statement valid for a
certain geometrical configuration continues to be valid if this
configuration is considered as a part of a larger configuration.
\enddefinition

\rm
Thus, the Alexandrian postulate means that an addition of any
subsidiary objects to the given geometrical configuration produces
no effects on it. The Alexandrian postulate is used, for instance, in
constructive proofs of geometrical theorems realized by the explicit
step-by-step construction of the objects, whose existence is stated by
the theorem. In case of a violation of the Alexandrian postulate the
process of constructive proof on each its step may use only configurations
compatible with preceeding ones, i.e. not changing the properties
that are the initial data of the theorem.

As an example of non-Alexandrian geometry one may consider the
Einstein geometry [7,2:App.A]. The kinematics and the process of
scattering of figures may be illustrated in the important example of
another non-Alexandrian geometry, the geometry of solitons [7,2:App.A].
All soliton geometries confirm the assumption that the breaking of the
Alexandrian postulate is connected with the interaction of geometric
figures, and in particular such interaction may be caused by the
nonlinear character of an evolution.

\define\tni{\operatorname{int}}
\define\txe{\operatorname{ext}}
Let us consider the process of scattering of figures in interpretational
geometry. As noted above, an interpretational figure is described by the
pair $(\Phi^{\tni},\Phi^{\txe})$, where $\Phi^{\tni}$ is an internal
image in the subjective space of the observer, and $\Phi^{txe}$ is
its drawing; $\Phi^{\tni}$ is the result of the interpretation of
$\Phi^{\txe}$. It is antural to assume that $\Phi^{\tni}$ depeneds on
$\Phi^{\txe}$ functionally: $\Phi^{\tni}_t=\Phi^{\tni}\left[
\Phi^{\txe}_{\tau\le t}\right]$ and, as a rule, nonlinearly; moreover,
if $\Phi^{\txe}$ is asymptotically free then $\Phi^{\tni}$ is also
asymptotically free. Consequently, the nontrivial scattering of
interpretational figures is observed even for the linear dynamics of
$\Phi^{\txe}$, so the interpretational figures are non-Alexandrian.

Thus, interpretational geometries realize a class of non-Alexandrian
geometries. The constructive proof in the interpretational geometry,
therefore, is in the explicit construction of the object, whose
existence is stated by the theorem, as a result of a step-by-step
process of constructing of intermediate configurations of interpretational
figures, whereas the correctness of proof is guaranteed on each step of
the proof by the claim for properties of the startpoint configuration
that are initial data of the theorem to be unchanged in the process of
interpretation. If the least is verified experimentally for each
concrete process then the proof has an experimentally mathematical
character being based on the incomplete induction; note that the similar
procedures are used in many self-educating systems of artificial
intelligence, in which possible combinations of steps are realized
(sometimes statistically) with subsequent verification of their
correctness (a method of the random seeking for solutions).

\subhead\bf 1.4. Informational aspects of interpretational geometries:
interpretational figures in the multi-user mode and IAVR-tele{\ae}sthesy
\endsubhead
This paragraph containes a brief exposition (without detailing) of
general results of the second part of the article [7], which were
not included into an appendix A to the article [2] and which treats
information aspects of interpretational geometries. The motivation
for such consideration lies in the remark of [7] that informatics
may be considered as a point of view on mathematical objectcs
complementary to geometric one (in view of the fundamental opposition
of ``logical'' and ``eidetical''). So it is convenient to reformulate
the main geometrical definitions in terms of informatics. Thus, the
introduced in the second part of the article [7] concepts of
AVR-photodosy and its formal grammar are a natural parallel to
concepts of anomalous virtual reality and transcending abstract model.
Note, however, that in view of methodological considerations
these concepts are considered now only purely formally and as
realizing the ``lower'' or ``external'' visual level of ``multifibred''
videocognitive semantic stream. Their deeper (but more than incomplete)
analysis was performed in the fourth paragraph of the article.

\definition{Definition 3A} An information transmission via
anomalous virtual reality by ``latent lights'' is called {\it
AVR-photodosy\/}; the system of algebraic structures of the
initial abstract model, which characterize AVR-photodosy under a
naturalization, is called the {\it formal grammar\/} of AVR-photodosy.
\enddefinition

Note that the concepts of AVR-photodosy and its formal grammar
are deeply related to one of anomalous color space because it is
namely using of such spaces allows to transmit diverse information
in various forms and as a sequence the investigation of problems
of information transmission via AVR, whose character depends on the
structure of an anomalous color space, is an important mathematical
problem (cf.[8]). A structure of AVR-photodosy is determined by its
formal grammar. As it will be shown in the next paragraph the
formal grammar of interpretational geometries may have a quantum
character and, therefore, the related AVR-photodosy obeys quantum
logic [9]; this fact undoubtly claims an intent attention as
theoretically obtained complex perceptive-cognitive analog of very
interesting and carefully investigated by scientific groups of
mathematicians, physicists, psychologists and neurophysiologists
(as theoretically as experimentally) purely cognitive processes,
which obey quantum logic. Note that the formal grammar of AVR-photodosy
in some concrete models of interpretational geometry was discussed in
the second part of the article [7] with the citing of numerous literature.

It is necessary to note that IAVR is polysemantic as a rule; it means that
a volume and a structure of information received by AVR-photodosy via it
depends on observer; so a natural problem is in the description of
informatics of interactive psychoinformation systems with more than one
observer, in particular, the correlation of different observations.
Such systems may be considered as realizing an interactive MISD
(Multiple\_Instruction--Single\_Data) architecture with parallel
interpretation processes for different observeres (this fact should be
regarded in the context of a remark on the quantum--logical character
of AVR-photodosy); on such way we encounter a phenomenon specific for
such systems but perphaps having a general meaning: namely, the
observation process for different observers generate an information
exchange between them.

\definition{Defintion 3B} AVR-photodosy via IAVR from an observer to
another is called the {\it IAVR-tele{\ae}sthesy\/}; if in the process
of IAVR-tele{\ae}sthesy AVR-photodosies from different observers do
not obey the superposition principle, we shall say that a {\it collective
effect\/} appears in the IAVR-tele{\ae}sthesy.
\enddefinition

Note that (1) the process of IAVR-tele{\ae}sthesy has a two-sided
character; observers entering IAVR-tele{\ae}sthetic communication
are simultaneously inductors (who send an informations) and
recepients (who receive it), moreover, a volume and a structure
of received information depends on a recepient as well as on an inductor;
(2) a collective effect in IAVR-tele{\ae}sthesy means that in IAVR the
inductors are not considered as independent, the transmitted information
is not a sum of informations sent by particular observers because the
partial information streams from each inductor participate in the
exchange interaction and form a specific information received by a
recepient. In this aspect it should be marked a relation of the
origin of IAVR-tele{\ae}sthesy to the fact that interactive
psychoinformation videosystems realize an interactive MISD architecture.

\definition{Definition 3C} An observer in the marginal AVR, to which
none observer can be related in the AVR realized on videocomputer, is
called {\it virtual\/}; and virtual observer, whose observation
process depend on several real observers, is called the {\it
collective virtual observer}.
\enddefinition

A presence of virtual observer means that a part of received information
is interpreted as an information sent by this really non-existing
observer. A presence of collective virtual observer is not
obligatory but common for interactive videosystems in a multi-user mode;
this fact should be also considered in the context of a remark that
such systems realize an interactive virtual MISD architecture with
parallel interpretation processes for different observers.

In the second part of the article [2] the exposed general concepts are
illustrated on a concrete model of interpretational geometry, in
particular an example of collective virtual observer is given that
clarify its meaning. Thus, it is mentioned that (1) only a part of
received information is interpreted as an information sent by collective
virtual observer (i.e. its presence does not demolish a presence of
real observers), (2) the process of intentional govern by collective
virtual observer is completely determined by interactions of real
observers (i.e. collective virtual observer is represented as
a specific unified state of real observers in the interactive
psychoinformation system in a multi-user mode), (3) collective
virtual observer enters an information exchange with real
observers being interpreted (at least, formally) as an independent
observer. In this case it is an important but completely unexplored
question on an interaction of individual real observers with
collective virtual observer, on a decomposition of the latter
on the non-correlated components ({\it quasisubjects}) and on their
interaction with each other.

In conclusion let us formulate a proposition proved at the end of the
second part of the article [7] by an explicit construction.

\proclaim{Proposition} {\it There exist models of interpretational
geometries in which there are interpretational figures observed only
in a multi-user mode.}
\endproclaim

It is rather interesting to explicate a meaning of mechanics of
observation of such figures in a context of general mathematical
game theory [10,11].

Unfortunately, in spite of all its advantages the described mechanism
of IAVR-te\-le\-{\ae}s\-the\-sy apparently does not allow to transmit
cognitive information directly. Namely to a description of an attempt to
adapt IAVR-tele{\ae}sthesy for a realization of the accelerated
nonverbal cognitive communication in integrated {\it videocognitive\/}
interactive systems the third and the fourth paragraphs of the article
are devoted.

\head\bf\S 2. Operator (quantum-field and stochastic) methods in the
theory of interactive dynamical videosystems and noncommutative
descriptive geometry\endhead

This paragraph is devoted to formalization and detailing of the exposed
in the previous one intuitively clear geometric picture and to
elaboration of the adequate algebraic and analytic methods. As a result
of realization of such intentions we shall have a possibility of
a more precise mathematical description of concrete models as
for further explication of general geometric aspects as for concrete
hardware and software elaboration for concrete interactive real-time
videosystems.

\subhead\bf 2.1. General operator aspects of interactive dynamical
videosystems and noncommutative descriptive geometry\endsubhead
There exist several widely used general ways to define an evolution of
images in the real-time interactive dynamical videosystems. Let us
expose some of them:
\roster
\item {\it Euler formulas\/} [12]:
$$\dot\Phi(t)=A(t,u,\dot u,\xi)\Phi(t),$$
where $u=u(t)\in\Bbb C\simeq\Bbb R^2$ is a current position of a sight
point (the screen is considered as a part of the complex plane
$\Bbb C\simeq\Bbb R^2$), $\dot u=\dot u(t)$ is the relative velocity
of its movement, $\xi=\xi(t)$ are additional parameters of an interactive
control, $\Phi=\Phi(t)$ is a set of continuously distributed chiral
(i.e. holomorphically-antiholomorphically decomposed, see [1]) visual
characteristics of an image (colors and overcolors), $A=A(t,u,\dot u,\xi)$
is a linear operator. The linear operator $A$ as a function on $u$ and
$dot u$ is called the {\it angular operator field}, the field $A$
(unbounded, in general) holomorphically depends on $u$ and $\dot u$ and
(weakly) continuous on $\xi$ in suitable (not obligatory metrizable)
general topology on the space of parameters of the interactive control
(e.g. biopotentials of EEG and ERG, dynamical parameters of a respiratory
rythm, etc., and also functional complexes of these magnitudes). The
dynamics of angular operator field on the variable $t$ may obey some
differential equation (e.g. Euler-Arnold equation [13]).
\item {\it Euler-Belavkin-Kolokoltsov formulas\/} [2]:
$$d\Phi(t,[\omega])=A(t,u,\dot u,\xi)\Phi(t)dt+\sum_{\alpha}B_\alpha(t,u,\dot
u,\xi)\Phi(t,[\omega])d\omega^{(\alpha)},$$
where $d\omega^{(\alpha)}$ is a set of stochastic differentials.
In practice, sometimes the fields $A(u,\dot u)$ and $B_\alpha(u,\dot u)$
may include (weak) nonlinearities. The dynamics of fields by the
variable $t$ may obey just as in the deterministic case to some
differential equation such as Euler-Arnold equation [13].
\item {\it Models with dynamical interctive screening\/} [1]:
In these models the collection of colors and overcolors $\Psi=\Psi(t)$ is
represented as
$$\Psi(t)=J(t,u,\dot u,\xi)\Phi(t),$$
where $\Phi=\Phi(t)$ obey Euler formulas or Euler-Belavkin-Kolokoltsov
formulas where\-as $J=J(t,u,\dot u,\xi)$ is a linear operator
(a projector with nontrivial kernel as a rule), which as a (holomorphic)
function on $u$ and $\dot u$ is called the {\it screening operator
field\/}; the screening operator field as well as the angular and other
operator fields accounted in the evolution equations for an image, is
(weakly) continuous on $\xi$ in a suitable (not obligatory metrizable
and, perhaps, specific for each operator field) general topology on
the space of parameters of the interactive control.
\item {\it Models with memory\/} [1]:
In this case the dynamics of colors and overcolors depends on the
hystory (e.g. is integrodifferential in time).
\endroster

Some concrete realizations of dynamics described above are rather
well investigated experimentally (for instance, the so-called
systems with partial dragging and masking -- see [1]).

The operator (quantum-field and stochastic) methods play an important
role in the analysis of all these models (on the stability of image, etc.)
(see e.g.[1,2,7,12,13]). As a rule the using of quantum-field methods is
based on the following supposition, which is called the {\it operator
algebra hypothesis}, namely that the coefficients of expansions of
operator fields, which enter the dynamical equations, by all variables
except $u$ being operator fields on $u$ form a closed operator algebra of
quantum field theory. So it is supposed that some magnitudes
characterizing an evolution of the system form an algebraic object, whose
definition is below. Note that in some specific classes of models the
hypothesis of operator algebra may be formally proved.

\definition{Definition 4A} The {\it QFT--operator algebra\/} (the {\it
operator algebra of quantum field theory\/}) is a linear space $H$
supplied with an operation $m_{\vec x}(\cdot,\cdot)$ depending on
the parameter $\vec x$ from $\Bbb R^n$ or $\Bbb C^n$, for which the
identity $m_{\vec x}(\cdot,m_{\vec y}(\cdot,\cdot))\!=\!m_{\vec y}(m_{\vec
x\!-\!\vec y}(\cdot,\cdot),\cdot)$ holds. Otherwords, QFT--operator
algebra is a pair $(H,t^k_{ij}(\vec x))$, where $H$ is a linear space
and $t^k_{ij}(\vec y)$ is a $H$--valued tensor field on $\Bbb R^n$ or
$\Bbb C^n$ such that $t^l_{im}(\vec x)t^m_{jk}(\vec y)\!=\!t^m_{ij}
(\vec y\!-\!\vec y)t^l_{mk}(\vec y)$. The field $t_{ij}^k(\vec x)$
realizes the operation $m_{\vec x}(\cdot,\cdot)$: $m_{\vec x}(e_i,e_j)=
t_{ij}^k(\vec x)e_k$, where $\{e_k\}$ is an arbitrary basis in $H$.
\enddefinition

Let us introduce the operators $l_{\vec x}(e_i)e_j=t^k_{ij}(\vec x)e_k$
(operators of the multiplication from the left in QFT--operator algebra)
then the following identities: $l_{\vec x}(e_i)l_{\vec y}(e_j)=t^k_{ij}(\vec
x-\vec y)l_{\vec y}(e_k)$ ({\it operator product expansion \/}) É $l_{\vec
x}(e_i)l_{\vec y}(e_j)=l_{\vec y}(l_{\vec x-\vec y}(e_i)e_j)$ ({\it
duality\/}) hold.

As a rule in the literature on mathematical physics the notation
$\varphi(\vec x)$ is used for operators $l_{\vec x}(\varphi)$ ($\varphi\in
H$). The quantities $\varphi(\vec x)$ are called operator fields.
In terms of operator fields the operator product expansions are written
as $$\varphi_1(\vec x)\varphi_2(\vec y)=F_\alpha(\vec x-\vec
y)\psi_{\alpha}(\vec y)$$
that means the decomposability of products of operator fields by
operator fields of QFT-operator algebra themselves. If a set of
concrete operator fields is closed under such decompositions then
they represent an abstract QFT-operator algebra.

If $dt_{ij}^k\equiv 0$ then QFT-operator algebra is an ordinary
associative algebra. An element $\varphi$ of QFT-operator algebra
is called the left divisor of zero if $l_{\vec x}(\varphi)\equiv 0$;
the unit in the QFT-operator algebra $H$ is an element $\boldkey 1$
such that $l_{\vec x}(\boldkey 1)\equiv\id$; the identity $\left.l_{\vec
x}(\varphi)\right|_{\vec x=0}=\varphi$ holds in any QFT-operator algebra
with unit and without left divisors of zero; if $V$ is the linear space
of left divisors of zero in the QFT-operator algebra $H$ then
$(\forall\varphi\in H)l_{\vec x}(\varphi)V\subseteq V$ and $H/V$ is
a QFT-operator algebra without left divisors of zero. There is defined
the operator $\bold L$ (the {\it operator of infinitesimal translations\/})
in any QFT-operator algebra with unit $\boldkey 1$: $\vec\bold L\varphi=
\left.\frac{d}{d\vec x}(l_{\vec x}(\varphi)\boldkey 1)\right|_{\vec x=0}$,
the operator of infinitesimal translations $\vec\bold L$ is a derivative
of QFT-operator algebra $H$ without left divisors of zero, i.e. for each
$\varphi$ from $H$ the identity $[\vec\bold L,l_{\vec x}(\varphi)]=l_{\vec
x}(\vec\bold L\varphi)$ holds; as a sequence $l_{\vec x}(\varphi)\boldkey
1=\exp(\vec x\cdot\vec\bold L)\varphi$. If $H$ is an arbitrary
QFT-operator algebra and $\vec\bold L$ is its derivative then the linear
space $\hat H=H\oplus\left<\boldkey 1\right>$ is supplied by the structure
of a QFT-operator algebra with unit $\boldkey 1$: $l_{\vec x}(\varphi)
\boldkey 1=\exp(\vec x\cdot\vec\bold L)\varphi$, $l_{\vec x}(\boldkey 1)=\id$.

Instead of QFT-operator algebras it is reasonable sometimes to consider
the related {\it local field algebras}, the ordinary associative algebras,
which are received from QFT-operator algebras by a renormalization
of the pointwise product of operator fields (note that as a rule tensor
fields $t_{ij}^k(\vec x)$ are singular at the point $\vec x=0$ and,
therefore, the pointwise product of operatorfields is either formally
infinite or indefinite). A procedure of renormalization was described in
the article [14]. For simplicity let us consider the specific case
when $\vec x=u$ is from the complex plane and operator fields are
meromorphic on $u$.

Besides the operator fields $\varphi(u)$ parametrized the elements
$\varphi$ of the space $H$ let us consider the expressions
$$\varphi(f)=\underset u=0\to\res\left\{f(u)\varphi(u)\frac{du}u\right\}=
\lim_{u\to 0}\left\{f(u)\varphi(u)-\text{singularities}\right\}.$$
The operator $\varphi(f)$ is corresponded to an element $\varphi$ of the
space $H$ and to a meromorphic function $f(u)$ (or to a meromorphic
1-form $f(u)\frac{du}u$). In view of operator product expansions
the product $\varphi_{\alpha}(f)\varphi_{\beta}(g)$ of two operators
$\varphi_\alpha(f)$ and $\varphi_{\beta}(g)$ is correctly defined and
admits a representation
$$\varphi_{\alpha}(f)\varphi_{\beta}(g)=
\varphi_{\gamma}(h^{\gamma}_{\alpha\beta}), \text{\ where \ }
h_{\alpha\beta}^{\gamma}(u)=\underset
v=u\to\res\left\{t_{\alpha\beta}^{\gamma}(v-u)
f(v)\frac{dv}v\right\}g(u).\tag{*}$$
This procedure is a renormalization of the pointwise product of
operator fields in QFT-operator algebra. The operator product expansions
are interpreted as a regularization of the pointwise product and functions
$f$ and $g$ as parameters, on which a result of renormalization depends.
An influence of change of functional parameters on the result of
renormalization (renorm-invariance) is described by formulas (*).

Operators $\varphi(f)$ are closed under their multiplication and form
an associative algebra $\frak A(H)$. This associative algebra is called
the local field algebra corresponded to the QFT-operator algebra $H$.
As a rule the noncommutative local field algebra $\frak A(H)$ for the
meromorphic QFT-operator algebra $H$ may be regarded as a structural ring
of some noncommutative manifold (noncommutative bundle over $\Bbb C\Bbb
P^1$ or noncommutative covering $\Bbb C\Bbb P^1$) (cf.[15-17]), and, thus,
to interpret the operator methods in the theory of the real-time
interactive dynamical videosystems as the {\it noncommutative
descriptive geometry\/} (cf.[13]).

\subhead\bf 2.2. Group theoretical and algebraic aspects of the interactive
dynamical videosystems\endsubhead As a rule it is reasonable to consider
those concrete models of interactive dynamical videosystems, which possess
some form of invariance with respect to geometrical transformations of
image or internal transformations of the color space. One of the simplest
form of a geometric invariance is one with respect to projective
transformations of $\Bbb C\Bbb P^1$, i.e. the group $\PSLTWO$ or its Lie
algebra $\sltwo$ though in a concrete realization really the models may be
considered with an invariance broken to translations and scaling
transformations (an example is the cutting off angular field in the
article [1]). The presence of projective invariance means the invariance
of evolution equations (Euler formulas, Euler-Belavkin-Kolokoltsov
formulas, etc.) that put some conditions on operators and operator
fields entering these equations. Projective invariance of operator fields
allows specify their form in a certain extent (in particular, to describe
their analytical dependence on $u$ and $\dot u$) what was done in [12].
On another hand the projective invariance of operator fields implies
the projective invariance of all constructed algebraic structures
(QFT-operator algebras and local field algebras). A systematical analysis
of the projectively invariant structures (operator algebras of quantum
projective field theory and local projective field algebras) was done
in the article [18]. Let us expose the necessary definitions and results
following [18,19].

\definition{Definition 4B {\rm [18,19]}} A QFT-operator algebra
$(H,t^k_{ij}(u); u\in\Bbb C)$ is called the {\it QPFT-operator algebra\/}
(the {\it operator algebra of quantum projective field theory\/}) if
(1) the space $H$ is a sum of Verma modules $V_{\alpha}$ over $\sltwo$
with extremal vectors $v_{\alpha}$ and extremal weights $h_{\alpha}$,
(2) $l_u(v_{\alpha})$ is a primary field of spin $h_{\alpha}$, i.e..
$[L_k,l_u(v_\alpha)]=(-u)^k(u\partial_u+(k+1)h_{\alpha})l_u(v_{\alpha})$,
where $L_k$ are $\sltwo$--generators ($[L_i,L_j]\!=\!(i\!-\!j)L_{i+j}$,
$i,j=-1,0,1$), (3) the descendant generation rule $L_{-1}l_u(f)=l_u(L_{-1}f)$
holds. A QFT-operator algebra $(H,w^k_{ij}(u);u\in\Bbb C)$ is called the
{\it derived QPFT-operator algebra} if the conditions (1) and (2) hold
together with the derived descendant generation rule:
$[L_{-1},l_u(f)]=l_u(L_{-1}f)$.
\enddefinition

As it was shown in the article [19] the categories of QPFT-operator
algebras and of derived QPFT-operator algebras are equivalent. An explicit
construction of their equivalence is presented there. As a sequence
QPFT-operator algebras and derived QPFT-operator algebras may be
regarded as different recordings of the same object and the most
convelient one may be used in a concrete situation.

It is necessary to note that $L=\ad(L_{-1})$ in QPFT-operator algebras with
unit whereas $L=L_{-1}$ in derived QPFT-operator algebras with unit.
Examples of QPFT-operator algebras were considered in [18,1,13].

In the structural theory of QPFT-operator algebras a crucial role is
played by one concrete QPFT-operator algebra, the algebra $\Vrt(\sltwo)$
of vertex operators for the Lie algebra $\sltwo$, whose construction
is contained in [18,19]. This algebra is realized in the skladen'
(= `unfolding' in Russian, here we use a direct transliteration)
of Verma modules over the Lie algebra $\sltwo$ and admits a strict
representation by operator fields in the model of Verma modules
over $\sltwo$. The meaning of this algebra is explicated by the following
theorem.

\proclaim{Theorem 1 {\rm (see [18])}} \it An arbitrary abstract
QPFT-operator algebra may be realized as a subalgebra in the algebra
$\Mat_n(\Vrt(\sltwo))$ of matrices with coefficients in $\Vrt(\sltwo)$
for some $n$ (finite if the QPFT-operator algebra contains a finite number
of primary fields and infinite otherwise).
\endproclaim

The algebra $\Vrt(\sltwo)$ of vertex operators for the Lie algebra $\sltwo$
has some additional features. Thus, for instance, its primary fields form
an infinite dimensional Zamolodchikov algebra [18].

A detailed information on this algebra, other QPFT-operator algebras,
local projective field algebras (LPFAs), their structural theory and
relations to classical algebraic structures may be found in [18,19].

Besides spatial symmetries the models of interactive dynamical videosystems
may possess internal symmetries, for instance, color ones [1]. If such
symmetries are described by the group $G$ then the related evolution
equations should be $G$--invariant and algebraic structures characterizing
the equations should possess $G$ as a symmetry group. The mechanisms
of that were carefully investigated in the article [13] (see also [1]).
In this case the main algebraic objects are the so-called projective
$G$--hypermultiplets, whose definition is given below.

\definition{Definition 4C {\rm [13]}} A QPFT--operator algebra (derived
QPFT--operator algebra) $(H,\mathbreak t^k_{ij}(u))$ is called the {\it
projective $G$--hypermultiplet} if the group $G$ acts in it by automorphisms,
otherwords, the space $H$ admits a structure of representation of the
group $G$, the operators of representation $T(g)$ commute with the
action of $\sltwo$ and $l_u(T(g)f)=T(g)l_u(f)T(g^{-1})$.
\enddefinition

The linear spaces of the extremal vectors of the fixed weight form
subrepresentations of the group $G$, which are called the {\it
multiplets\/} of the projective $G$--hypermultiplet. Projective
$G$--hypermultiplets with $G\supseteq\SU(3)$ describe the algebraic
structure of a color space in projectively invariant IAVR [1]. Some
concrete projective $G$--hypermultiplets (canonical and
hypercanonical ones as well as their specific cases, the $q_R$--conformal
field theories) were investigated in details in the articles [13,1,18].
Note also that directly of after a slight modification many methods of a
well-elaborated quantum conformal field theory are applicable to these
concrete projective $G$--hypermultiplets [13,18] (in particular,
one may adapt the Sugawara construction, Fubini-Veneziano fields and vertex
constructions, Virasoro master equations, modular functor, etc. to
them).

\head\bf\S 3. Organization of the integrated real-time interactive
videocognitive systems\endhead

This paragraph is devoted to general principles of an organization
of integrated real-time interactive videocognitive systems.
Integrated systems consist of two simultaneously working subsystems,
one of which is artificial (an interactive dynamical videosystem as a
rule) and another is natural (sensorial, respiratory, cognitive or
complex). It is necessary to mention that the most of all real
videosystems are implicitely or explicitely integrated videocognitive
ones, so it is a strong interest namely to the integrated interactive
systems.

However, before to begin an investigation of integrated interactive
system it is necessary to give an analysis of natural interactive
systems to understand the laws of their working as parts of integrated
systems. It is reasonable to begin from the simplest sensorial--type
systems, then to transit to the related integrated systems (videosensorial
and videorespiratory) and only then to concentrate an attention on
the properly videocognitive systems.

\subhead\bf 3.1. Virtualization as method of the description of a
class of physical interactive information systems [18,20]\endsubhead
Let us expose following [18,20] the general principles of
virtualization of natural physical interactive systems as a method of
their kinematical description.

\definition{Definition 5 {\rm [20]}} The {\it image of a natural
interactive system by artificial interactive videosystem\/} is
the fixing of an algorithm of the construction of a dynamical image of
any interactive process in the natural system by use of the intentional
anomalous virtual reality realized by the computer--graphic interface of
the artificial interactive videosystem. The {\it virtualization of
a natural interactive system\/} is the constructing of its image by
the artificial interactive videosystem using some set of experimental data
on the system in the stationary regime. The initial natural interactive
system will be called the {\it realisation\/} of the artificial interactive
system.
\enddefinition

Otherwords, virtualiation of a natural interactive system allows to
construct a dynamical image of any arbitrary (not obligatory
self-oscillating) interactive process by some artificial interactive
system using only haracteristics of a certain finite set of
self-oscillating interactive processes. A choice of the term
``virtualization'' is motivated by the fact that a dynamical image of
the interactive process is an anomalous virtual reality in a wide sense.

In view of the decomposition of the natural interactive systems of
our interest onto an active and a passive components (the active and
the passive agents) the characteristics of the stationary self-oscillating
process are divided on the graphical data on the dynamical state of the
passive agent and the amplitude-frequency characteristics of the active
agent. Virtualization of a natural interactive system is reduced to an
algorithm of the {\it secondary synthesis\/} [20] of graphical data on
a dynamical state of the passive agent in any interactive process by
the fixed collection of such data for some set of the stationary
self-oscillating processes as well as of the corresponding
amplitude-frequency characteristics of the active agent in such processes;
the principal structurally algorithmic scheme of virtualization of
natural interactive systems based on the using of the secondary synthesis
is exposed in the article [20].

{\it The criterion of the adequacy of an image and of the correctness of
virtualization {\rm [20].}} A virtualization of the natural interactive
system is correct if the received image of this system by some
artificial interactive videosystem is adequate to the initial system,
i.e. under the simultaneous working of both artificial and
natural interactive systems with the same active agent the
reproducing of an image of certain interactive process (under the
certain state and decision of the active agent) allows to provoke its
controlled performance in the natural interactive system.

Certainly, the image may be constructed one time on the basis
of the experimental data on the working of the natural interactive
system with an active agent different from that acts in the
reproducing of the image. Note also that under the reproducing of the
image a simultaneous collective working of two different interactive
(artificial and natural) systems with the same subject is supposed.
Thus, the main difference of the virtualization from the mathematical
model is in the fact that the least is a {\it pure descriptor\/} of
the physical process whereas the first is its {\it descriptor--constructor}.

Let us formulate some phenomenological principles, whose fulfilment is
apparently necessary for a correctness of virtualization:\newline
-- {\it The principle of isostructural ideographicity of virtualization
{\rm [20]:}} Virtualization of a natural interactive system should be
performed in a manner that the obtained image and the initial system
have the same structures;\newline
-- {\it The principle of dynamical equivalency of intentions {\rm [20]:}}
Virtualization of a natural interactive system should be performed in a
manner that the parameters characterizing the intensity of intension
of a subject on an object coincide for its and its image during the
dynamical subject-object interaction. One may consider a correlation
of the behavioral reactions of a subject and dynamical characteristics of
an object as such parameter.
\pagebreak

\subhead\bf 3.2. Examples of an organization of the integrated
interactive systems and perspectives for their development [20]
\endsubhead
The interactive systems may be divided on the following classes:
\roster
\item"--" {\it By the origin:\/} on artificial, natural and integrated.
\item"--" {\it By the type of interface\/} on videosystems,
audiosystems, general sensorial (including tactile), respiratory and
complex (audiovideosystems, videosensorial, videorespiratory, etc.).
\item"--" {\it By the goals:\/} on the systems of management,
the systems of observation, analysis and accumulation of information,
the systems for self-regulation and autotraining, the communication
systems.
\item"--" {\it By the number of users:\/} on the systems in one- and
multi-user modes.
\endroster

The most interesting systems (among many others) are the following
types of the interactive systems: (1) artificial videosystems of
the ``automatical arrangement of perception'', including multi-user
ones based on the secondary image synthesis (SIS) [20], (2)
the systems, which include the preceeding ones as a subsystems, the
integrated videosensorial systems of an interactive vision, (3)
the integrated videorespiratory systems for self-regulation,
autotraining and cognitive stimulation, (4) integrated multi-user
videocognitive systems for an accelerated nonverbal communication.

The most perspective in the SIS systems [20], which play a key role
in the projecting of many integrated systems, is apparently an
investigation of intellectual SIS systems with a dynamical interactive
tuning (as well as an eleboration of portative videosystems of
synchronous SIS and complex multi-user systems). The SIS systems with
dynamical interactive tuning may be considered as systems of
the ``automatical arrangement of perception'' in the context of an
activity on the computer systems of the interactive ``automatical
painting'' as well as of investigations of the synthetical perception.
It is reasonable to study octonionic anomalous 3D stereosynthesis [1],
chiral dissimmetry of the visual analyzer in interactive processes
and its iinfluence on the stereosynthesis. It is important to
investigate the simultaneous working of natural interactive systems and
their virtualizations for an elaboration of the integrated interactive
systems realizing an {\it ``integrated reality''\/} as an alternative
to the systems of ``virtual reality''. To the integrated systems,
which include the artificial interactive videosystems for the automatical
arrangement of perception as parts, one should attribute the integrated
videosensorial systems of the interactive vision with a natural sensorial
subsystem and the integrated videorespiratory interactive systems
(i.e. interactive videosystems with a respiratory modulation or, in
general, a transformation of visual interactive processes) for the
psychophysiological autotraining (as passively relaxational as actively
dynamical).

\subhead\bf 3.3. An organization of the integrated real-time videocognitive
systems for the accelerated nonverbal cognitive communication\endsubhead
Some cognitive aspects of the interactive computer graphics in general
were considered in the monograph [21]. In this article we are
interested in the cognitive aspects of the communication real-time
interactive videosystems (as well as their integrated videosensorial
counterparts).

As it was marked above the direct using of IAVR--tele{\ae}sthesy does
not apparently allow to realize a cognitive information transfer.
On another hand integrated videocognitive communication interactive
systems can ``inherit'' the useful properties of their subsystems
realizing IAVR--tele{\ae}sthesy but also can transfer a complex
cognitive information. Thus, an approach to elaboration of such ``derived''
integrated systems for an accelerated nonverbal cognitive communication
based on application of IAVR--tele{\ae}sthesy (see above) as a mechanism
of the preliminary recognition and the interactive {\it self-tuning\/}
of communications ``keys'' of subjects as well as their dynamical
synchronization should provide the users by the channelwise interactive
establishing of the communication ``keys'' and, therefore, the principal
noninterpretability (nondecipherability) of the communication from the
outside. The realizability of such scheme is based on several
circumstances: (1) the presence of interpretational figures observed
only in the multi-user mode together with other features of the multi-user
mode in interpretational geometry discussed above (e.g. the polysemanticity
and quantum logical character of communication, which allows to use some
ideas and methods of quantum cryptography [22-25]), (2) the ability of
interpretational figures to be pointers on the videocognitive objects on
another nature, which are dynamically reconstructed in real time by users
during the communication. Such objects will be called the {\it dynamically
reconstructed objects of experimental mathematics\/} of briefly {\it
``droems''}. So the stream of visual information between users, which
was analyzed in \S1.4, organizes a more complex stream of videocognitive
information between them (note that though the transmitted information
is cognitive the process of its transmission has an interactively
controlled and partially subconscious character so that the two-sided
stream of the interactive videoinformation may be considered as a
psychosemantical context of the cognitive information exchange).
The initial interpretational geometry (more precisely, the interactive
videosystem, its ``bearer'') is a virtualization (a descriptor-constructor)
of the realized cognitive interactive system. To a discussion of the
mathematical aspects of the theme, namely, to droems and their dynamical
reconstruction we shall transit in the following paragraph.

\head\bf\S 4. Droems and their dynamical reconstruction\endhead

This paragraph is devoted to droems and their dynamical reconstruction.
Though in general this process as well as a general nature of droems
are not completely clear even on a conceptual level (though droems are
apparently related to the so-called dynamical simulacres of notions
[20:note 5]) some individual examples may be analyzed rather formally,
that makes clear some general mechanisms. The example, which will be
treated below, is related to the infinite dimensional geometry and may
be characterized as an answer on the question whether it is possible to
observe infinite dimensional objects, otherwords, whether the {\it
descriptive infinite dimensional geometry\/} is possible and how. As
we shall see the using of the interactive videocognitive systems,
droems and their dynamical reconstruction solves this question
positively in principle in some sense that {\it allows to use the
methods of experimental (computer) mathematics for an investigation
of infinite dimensional geometrical objects}. Note that some aspects
of the infinite dimensional descriptive geometry were theoretically
discussed in [26].

It is convenient to discuss two technical questions (see 4.1. and
4.2. below) to the example itself (see 4.3., 4.4. below).

\subhead 4.1. Organization of the cyberspace [7,2:App.A]\endsubhead
For simplicity let our image dynamics be defined by the Euler formulas,
which are coupled with some explicit dependence of angular fields on
time, so that their collection is projective-invariant in scope.
The cyberspace consists of the image space $V_I$ with the fundamental
length (the step of lattice) $\Delta_I$ and the observation space $V_O$
with the fundamental length $\Delta_O$; the space $V_I$ is one for
images whereas $V_O$ is used for data on the eye motions; it is naturally
to claim that $\Delta_I\gg\Delta_O$. The Euler formulas may be written as
$\dot\Phi_t=A_{t,\xi}(u,\dot u)\Phi_t$, we shall suppose that the
components of a decomposition of the angular field $A_{t,\xi}(u,\dot u)$
by $\dot u$ generate a QPFT--operator algebra of $q_R$--conformal field
theory [18], i.e. they are the $\sltwo$--primary fields in the Verma
module $V_h$ over the Lie algebra $\sltwo$. The angular field
$A_{t,\xi}(u,\dot u)$ may be approximately represented as $M_1(t,\xi)\dot
uV_1(u)+M_2(t,\xi)\dot u^2V_2(u)+\dots+M_n(t,\xi)\dot u^nV_n(u)$, where
the magnitudes $M_i(t,\xi)$ realize the explicit dependence of the
angular field on time and nongeometric dynamical parameters such as
biopotentials of EEG or data on the respiratory rythm, and $V_i(u)$ are
$\sltwo$--primary fields of spin $i$ in the Verma module $V_h$ over the
Lie algebra $\sltwo$. The explicit formulas for these fields are
written, for instance, in [16], as well as their discrete (lattice)
counterparts, which are used in practice (in this case $n$ is equal
either to 2 or 3).

\define\cut{\operatorname{cut}}
\subhead 4.2. Cutting off the angular field [1]\endsubhead
From the practical point of view it is reasonable to consider
the cutting off $A^{\cut}(u,\dot u)$ the angular field $A(u,\dot u)$
in its regular part, which does not contain the degrees of $u$ more
than $N$. It is supposed that $A^{\cut}(u,\dot u)$ is translation and
scaling invariant whereas the dilatation invariance is broken. For
instance, the components of the cut-off $q_R$--affine current $J^{\cut}(u)$
($\sltwo$--primary field of spin 1 in Verma module $V_h$ over the Lie
algebra $\sltwo$) are dfined by the operators $J_k^{\cut}=J_k=\partial_z^k$,
$J_{-k}^{\cut}=z^k\Delta_+^kP(z\partial_z)$ ($\Delta_+f(x)=f(x+1)-f(x)$),
where $P(z\partial_z)$ is a polynomial of a degree $N$, for instance,
such that $P(z\partial_z)z^i=\frac1{2h+i}z^i$, $i\le N$. The Verma module
$V_h$ is realized in the space of polynomials of a variable $z$,
$\sltwo$--generators have the form $L_1=(z\partial_z+1h)\partial_z$,
$L_0=z\partial_z+h$, $L_{-1}=z$. The polynomial $P(z\partial_z)$ uniquely
determines an operator $L_1^{\cut}$ such that $[L_1^{\cut},J_{-1}^{\cut}]=1$,
$[L_1^{\cut},L_0]=L_1^{\cut}$; $L_1^{\cut}$ is the cut-off dilatation
operator: $L_1^{\cut}=zP^{-1}(z\partial_z)$. The operators $L_1^{\cut}$,
$L_0$, $L_{-1}$ form the so-called nonlinear $\operatorname{sl}_2$ [27]
with relations
$$[L_0,L_{-1}]=L_{-1},\quad [L_1^{\cut},L_0]=L_1^{\cut},\quad
[L_1^{\cut},L_{-1}]=h(L_0),$$
where $h(x)=\frac1{P(x+1)}-\frac1{P(x)}$. This nonlinear
$\operatorname{sl}_2$ describes a breaking of projective invariance
under the cutting off procedure.

In particular cases the cutting off procedure at $N=1$ and
$A(u,\dot u)=J(u)\dot u$ realizes an interactive videosystem with
a partial dragging and masking, i.e.
$$\Phi(x)=\Phi_u(x)=f(|x-u|)\Phi_0(x-\gamma u),$$
where $\gamma$ is the dragging coefficient, $f$ is the masking function,
$u=u(t)$ is the sight point position. Videodata may be multifibred with
the dragging coefficient and the masking function specific for each fiber.

A transition to the discrete (lattice) version after the cutting off
does not provoke any problems.

\subhead 4.3. Infinite dimensional dynamical symmetries of the
interactvely controlled videosystems\endsubhead
The infinite dimensional dynamical symmetries of the interactively
controlled videosystems, whose evolution is described by the Euler
formulas or Euler-Belavkin-Kolokoltsov formulas (perhaps coupled with
Euler-Arnold equations) with operator fields generating a QPFT--operator
algebra of $q_R$--conformal field theory may be constructed in one of
three equivalent ways.

{\it Mode 1\/} [18]. Generators of the infinite dimensional dynamical
symmetries are just the $\sltwo$--tensor operators in Verma modules $V_h$
over the Lie algebra $\sltwo$, which are transformed under an action
$\sltwo$ as holomorphic $n$--differentials in the unit complex disk $D_+$
($n\in\Bbb Z_-$), i.e. as $m$--polyvector fields ($m\in\Bbb Z_+$).
Therefore, the generating functions for generators of the infinite
dimensional dynamical symmetries are $\sltwo$--primary operator fields
in $V_h$ of spin $m$.

{\it Mode 2}. The action of the Lie algebra $\sltwo$ in its Verma module
$V_h$ may be extended to the representation $T$ of the Lie algebra $W_1$
of formal vector fields on a line [28]. Namely, if the Verma module $V_h$
is realized in the space of polynomials of a complex variable $z$ and
$\sltwo$--generators have the form $L_{-1}=z$, $L_0=z\partial_z+h$,
$L_1=z\partial^2_z+2h\partial_z$ then other generators of the Lie algebra
$W_1$ are defined by the operators $L_k=z\partial_z^{k+1}+(k+1)h\partial_z^k$
($k\ge 2$). If the Verma module $V_h$ is unitarizable then the generators
of the infinite dimensional dynamical symmetries corresponded to the
$\sltwo$--primary field of spin 2 are represented in the form $T(X)$ or
$T^*(X)$, where $X\in W_1$. In the nonunitarizable case one should use
an analytical continuation on the parameter $h$. To receive generators
of the infinite dimensional dynamical symmetries related to the
$\sltwo$--primary field of spin 1 it is necessary to extend the action of
the Lie algebra $W_1$ in the Verma module $V_h$ to the representation
$\tilde T$ of the semi-direct sum of this algebra and the abelian Lie
algebra $\Bbb C[z]$ in $V_h$. This representation being reduced onto
$\Bbb C[z]$ is a representation of this algebra not only as abelian Lie
algebra but also as a commutative associative algebra, its generator $z$
is mapped to the operator $\partial_z$. The generators of the infinite
dimensional dynamical symmetries related to the $\sltwo$--primary field
of spin 1 has the form $\tilde T(X)$ or $\tilde T^*(X)$, where $X\in\Bbb
C[z]$. An analogous slightly more complicated construction allows to
receive other infinite dimensional dynamical symmetries.

{\it Mode 3}. Let us extend the representation of the Lie algebra $\sltwo$
in its Verma module $V_h$ to a representation $T$ of the Lie algebra
$\Vect^{\Bbb C}(S^1)$ of smooth $\Bbb C$--valued vector fields on a circle
$S^1$ (more precisely, of the $\Bbb Z$--graded Witt algebra of polynomial
vector fields) in the module of functional dimension 1. Let us denote by
$P$ the natural $\sltwo$--invariant projector of the space $\End(V(h))$
onto the space $\End(V_h)$. The infinite dimensional dynamical symmetries
corresponded to the $\sltwo$--primary field of spin2 have the form
$P(T(X))$ ($X\in\Vect^{\Bbb C}(S^1)$). To receive other infinite
dimensional dynamical symmetries it is necessary to consider the Lie
algebra $\DOP^{\Bbb C}_{[\cdot,\cdot]}(S^1)$ of all differential
operators instead of the Lie algebra $\Vect^{\Bbb C}(S^1)$ of all vector
fields.

An algebraic structure of these infinite dimensional dynamical symmetries
was unraveled in the articles [29,30]. Let us give the final formulation
of results.

\proclaim{Theorem 2A} \it The infinite dimensional dynamical symmetries
corresponded to the $\sltwo$--primary field of spin 2 form a
$\HS$--projective representation of the Lie algebra $\Vect^{\Bbb C}(S^1)$
in the unitarizable Verma module $V_h$ (i.e. a representation up to
Hilbert-Schmidt operators) as well as an asymptotic representation of
this algebra ``$mod\ O(\hbar)$'' (in sense of [31]; $\hbar=h-\frac12$).
All collection of the infinite dimensional dynamical symmetries form
a $\HS$--projective and asymptotical ``$mod\ O(\hbar)$'' representations
of the Lie algebra $\DOP^{\Bbb C}_{[\cdot,\cdot]}(S^1)$.
\endproclaim

The statement of the theorem can be easily derived from the third
mode to define the infinite dimensional dynamical symmetries.

A part of the infinite dimensional dynamical symmetries may be
``globalized'' [30]. Let us formulate the final result.

\proclaim{Theorem 2B} \it The infinite dimensional dynamical symmetries
corresponded to the $\sltwo$--primary fields of spin 1 and 2 are
exponentiated to a projective $\HS$--pseudorepresentation [30] of
a semi-direct product of the group $\Diff_+(S^1)$ of diffeomorphisms of
a circle and the loop group $\Map(S^1,U(1))$, which is also an asymptotic
representation ``$mod\ O(\hbar)$''.
\endproclaim

The theorem 2B admits an analog for the infinite dimensional dynamical
symmetries for dynamics in arbitrary canonical projective
$G$--hypermultiplets [13,1]. For that one should change the abelian
group $U(1)$ to the group $G$.

Let us consider an approximation of the angular field with $n=2$
(see 4.1.). In this case the angular field is represented in the form
of generators of the infinite dimensional dynamical symmetries with
coefficients depending on the control parameters. hence, the dynamics is
integrated up to Hilbert-Schmidt operators or asymptotically
``$mod\ O(\hbar)$'' and is defined by the interactively controlled
group element of the semi-direct product of the group of diffeomorphisms
of a circle and the loop group.

\subhead 4.4. Infinite dimensional droems and their dynamical
reconstruction\endsubhead
From the results of 4.3. it follows that the interpretational figures
in the models defined by canonical projective $G$--hypermultiplets may
be pointers to the infinite dimensional droems realized by use of
geometrical objects related to the groups of diffeomorphisms of a
circle and the loop groups (see [26,32-35,17] and refs wherein).
The process of a dynamical reconstruction is in the restoring of
the infinite dimensional object by an interactive process in the
dynamical videosystem. The infinite dimensional object itself may be
as interpretational as static (compilational). For instance,
interpretational figures may be pointers to static ``image'' on the
``infinite dimensional screen'' -- the space of the universal deformation
of a complex disc [33,34]. Droems may be looked for among numerous
(sometimes, rather exotic) infinite dimensional objects of the
geometrical theory of second quantized strings [35] as it was marked
in the third part of the article [35]; so it is reasonable to speak on
the {\it descriptive string geometry\/} in the context of a general
mathematical formalism for string theory (see e.g.[36]).

Note that a transition to the lattice version should apparently
produce quantum analogs of infinite dimensional Lie groups and Lie
algebras (cf.[37]), however, the geometrical consequences of the
cutting off procedure (see 4.2. above) are not known.

An importance of the infinite dimensional geometry was stressed in [26];
this article may be regarded as a development of the thesis of the
naturality of a language of infinite dimensional geometry formulated in [26].
Note that droems (in particular, infinite dimensional) and their dynamical
reconstruction besides the described applied problem of an organization
of accelerated nonverbal cognitive computer and telecommunications
being very important for experimental mathematics and, therefore, for
the whole complex of mathematical sciences are of interest for
theoretical mathematical psychology, structural linguistics and linguistic
psychology in the context as of investigations of nonverbal cognitive
communications in various external conditions (e.g. with stimulations),
which became actual recently, as for more traditional themes such as
investigations of origins and development of verbal communications,
early stages of speech and education processes. So it is of a special
interest to consider dynamical reconstruction of droems in the multi-user
mode (cf.1.4), for instance, in the interactive videocognitive games
(the games with an interactive control, see 1.1.).

\Refs
\roster
\item" [1]" Juriev D.V., Octonions and binocular mobilevision [in
Russian]. Fundam.Prikl.Matem. 1998, to appear [Draft English e-version: 
hep-th/9401047 (1994)].
\item" [2]" Juriev D.V., Belavkin-Kolokoltsov watch-dog effects in
interactively controlled stochastic dinamical videosistems. 
Theor.Math.Phys. 106 (1996) 276-290.
\item" [3]" Kalawsky R.S., The science of virtual reality and virtual
environments. Addison-Wesley, 1993.
\item" [4]" Virtual reality: applications and explorations. Ed.A.Wexelblat.
Acad.Publ., Boston, 1993.
\item" [5]" Burdea G., Coiffet Ph., Virtual reality technology. J.Wiley \&\
Sons, 1994.
\item" [6]" Forman N., Wilson P., Using of virtual reality for
psychological investigations [in Russian]. Psihol.Zhurn. 17(2) (1996) 
64-79.
\item" [7]" Juriev D., Visualizing 2D quantum field theory: geometry and
infomatics of mobilevision: Report RCMPI-96/02 (1996) [Draft e-version:
hep-th/9401067+hep-th/9404137 (1994)].
\item" [8]" Saaty T.L., Speculating on the future of Mathematics.
Appl.Math.Lett. 1 (1988) 79-82.
\item" [9]" Beltrametti E.G., Cassinelli G., The logic of quantum mechanics.
Encycl.Math.Appl.15, Ad\-di\-son-Wesley Publ., London, 1981.
\item"[10]" Owen G., Game theory. Saunders, Philadelphia, 1968.
\item"[11]" Vorob'ev N.N., Game theory [in Russian]. Leningrad, 1985.
\item"[12]" Juriev D.V., Quantum projective field theory:
quantum-field analogs of the Euler formulas. Theor.Math.Phys. 92
(1992) 814-816.
\item"[13]" Juriev D.V., Quantum projective field theory: quantum field
analogs of the Euler-Arnold equations in projective $G$--hypermultiplets.
Theor.Math.Phys. 98 (1994) 147-161.
\item"[14]" Juriev D.V., QPFT--operator algebras and commutative 
exterior differential calculus. Theor.Math. Phys. 93 (1992) 1101-1105.
\item"[15]" Witten E., Non-commutative geometry and string field theory.
Nucl.Phys. B268 (1986) 253-291.
\item"[16]" Witten E., Quantum field theory, grassmannians and algebraic
curves. Commun.Math.Phys. 113 (1988) 529-600.
\item"[17]" Juriev D.V., Quantum conformal field theory as infinite
dimensional geometry. Russian Math.Sur\-veys 46(4) (1991) 135-163.
\item"[18]" Juriev D.V., Complex projective geometry and quantum
projective field theory. Theor.Math.Phys, 101 (1994) 1387-1403.
\item"[19]" Bychkov S.A., Juriev D.V., Three algebraic structures
of quantum projective ($\sltwo$--invariant) field theory. Theor.Math.Phys.
97 (1993) 1333-1339.
\item"[20]" Juriev D.V., On the description of a class of physical
interactive information systems [in Russian]: Report RCMPI-96/05 (1996) 
[e-version: mp\_arc/96-459 (1996)].
\item"[21]" Zenkin A.A., Cognitive computer graphics [in Russian]. Moscow, 
Nauka, 1991.
\item"[22]" Wiesner S., Conjugate coding. SIGACT News 15(1) (1983) 78-88.
\item"[23]" Wiedemann D., Quantum cryptography. SIGACT News 18(2) (1989)
28-30.
\item"[24]" Bennett C.H., Brassard G., The dawn of a new era for quantum
cryptography: the experimental prototype is working. SIGACT News 20(2)
(1989) 78-82.
\item"[25]" Bennett C.H., Brassard G., Cr\'epeau C., Jozsa R., Peres A.,
Wootwers W.K., Teleporting and unknown quantum state via dual classical and
EPR channels. Phys.Rev.Lett. 70 (1993) 1895-1899.
\item"[26]" Juriev D., The vocabulary of geometry and harmonic analysis on the
infinite--dimensional manifold $\Diff_+(S^1)/S^1$. Adv.Soviet Math. 2 (1991)
233-247.
\item"[27]" Ro\v cek M., Representation theory of the nonlinear $\SU(2)$
algebra. Phys.Lett. B255 (1991) 554-557.
\item"[28]" Fuks D.B., Cohomology of infinite dimensional Lie algebras
[in Russian]. Moscow, Nauka, 1983.
\item"[29]" Juriev D., Topics in hidden symmetries. V. E-print:
funct-an/9611003.
\item"[30]" Juriev D., On the infinite-dimensional hidden symmetries. I-III.
E-prints: funct-an/9612004, funct-an/9701009, funct-an/9702002.
\item"[31]" Karasev M.V., Maslov V.P., Nonlinear Poisson brackets.
Geometry and quantization. Amer.Math. Soc., Providence, R.I., 1993.
\item"[32]" Juriev D.V., Non-Euclidean geometry of mirrors and 
pre-quantization on the homogeneous K\"ahler manifold
$M=\Diff_+(S^1)/\Rot(S^1)$ [in Russian]. Uspekhi Matem.Nauk. 43(2) (1988)
187-188.
\item"[33]" Juriev D.V., A model of Verma modules over the Virasoro
algebra. St.Petersburg Math.J. 2 (1991) 401-417.
\item"[34]" Juriev D., Infinite--dimensional geometry of the universal
deformation of the complex disk. Russian J.Math.Phys. 2 (1994) 111-121.
\item"[35]" Juriev D., Infinite dimensional geometry and quantum field theory
of strings. I-III. Alg.Groups Geom. 11 (1994) 145-179 [e-version:
hep-th/9403068 (1994)]; Russian J.Math.Phys. 4 (1996) 287-314;
J.Geom.Phys. 16 (1995) 275-300.
\item"[36]" Green M., Schwarz J, Witten E., Superstring theory.
Cambridge Univ.Press, Cambridge, 1987.
\item"[37]" Reshetikhin N.Yu., Semenov-Tian-Shansky M.A., Central extensions
of quantum current groups. Lett.Math.Phys. 19 (1990) 133-142.
\endroster
\endRefs
\newpage

\cyr UDK: \rm 519.689.6+519.76+519.723.4\qquad\qquad\qquad\qquad
$\boxed{\boxed{\aligned
&\text{\eightpoint"Thalassa Aitheria" Reports}\\
&\text{\eightpoint RCMPI-96/05$^+$}\endaligned}}$\newline
\ \newline
\ \newline
\document
\centerline{\cyb DROE1MY: E1KSPERIMENTALP1NAYa MATEMATIKA,
INFORMATIKA I}
\centerline{\cyb BESKONEChNOMERNAYa GEOMETRIYa}
\ \newline
\centerline{\cyr D.V.Yurp1ev}
\ \newline
\ \newline
\centerline{\cyie Tsentr matematicheskoe0 fiziki i informatiki}
\centerline{\cyie ``Talassa E1teriya''}
\centerline{\it E-mail:\/ \rm denis\@juriev.msk.ru}
\ \newline
\ \newline
\centerline{cs.HC/9809119}
\ \newline
\ \newline
\ \newline

\cyr Dannaya rabota, adresovannaya kak matematikam-teoretikam,
spe\-tsi\-a\-li\-zi\-ru\-yu\-shchimsya v geometrii, algebre,
funktsionalp1nom analize, teorii dinamicheskih i upravlyaemyh sistem,
interesuyushchimsya sovremennymi prilozheniyami ih distsiplin
k zadacham informatsionnyh tehnologie0 (organizatsiya interaktivnyh
dinamicheskih videosistem realp1nogo vremeni dlya uskorennyh
kompp1yuternyh i telekommunikatsie0), tak i prikladnikam,
zanimayushchimsya razrabotkoe0 ukazannyh sistem s aktsentom na
matematicheskie metody interaktivnoe0 kompp1yuternoe0 grafiki
v realp1nom vremeni, posvyashchena analizu opredelennogo, po-vidimomu,
dostatochno perspektivnogo, s odnoe0 storony, a s drugoe0 --
nebezynteresnogo s tochki zreniya chistoe0 (kak e1ksperimentalp1noe0,
tak i teoreticheskoe0) matematiki podhoda k resheniyu odnoe0
problemy informatsionnyh tehnologie0.

Osnovnoe0 trudnostp1yu, prepyat\-stvuyushchee0 dalp1nee0shemu
razvitiyu in\-te\-rak\-tiv\-nyh sistem realp1nogo vremeni dlya
uskorennyh kompp1yuternyh i te\-le\-kom\-mu\-ni\-ka\-tsie0, yavlyaet\-sya
sushchestvennaya ogranichennostp1 obp2ema informatsii, peredavaemoe0
za edinitsu vremeni. Esli dlya neinteraktivnyh sistem i
interaktivnyh sistem uslovnogo vremeni, prednaznachennyh dlya
osushchestvleniya uskorennyh kommunikatsie0, e1ta trudnostp1 bolee
ili menee preodolevaet\-sya blagodarya is\-polp1\-zo\-va\-niyu razlichnyh
priemov szhatiya i kodirovaniya pri odnovremennom
sovershenstvovanii materialp1noe0 (tehnicheskoe0) bazy, to dlya
interaktivnyh sistem realp1nogo vremeni pri naivnom podhode ona
stanovit\-sya pochti nepreodolimoe0. Krome togo, primenenie
interaktivnyh sistem realp1nogo vremeni dlya organizatsii uskorennyh
kommunikatsie0 na bolp1shih rasstoyaniyah porozhdaet dopolnitelp1nuyu
trudnostp1, svyazannuyu s vyzvannoe0 relyativist\-skimi e1ffektami
nevozmozhnostp1yu sinhronizatsii ritmov subp2ektov kommunikatsii;
tak, pri ras\-sto\-ya\-nii v $N$ tysyach kilometrov pogreshnostp1 v
sinhronizatsii teoreticheski ne menp1\-she, chem $N/300$ sekund, a
prakticheski nachitelp1no bolp1she.

Perechislennye trudnosti privodyat k neobhodimosti inoe0
organizatsii informatsii i ee peredachi v interaktivnyh sistemah
realp1nogo vremeni dlya uskorennyh kommunikatsie0. Odnoe0 iz takih
vozmozhnostee0, kstati ispolp1zuemoe0 v sistemah obychnyh 
kommunikatsie0, yavlyaet\-sya peredacha po kanalam svyazi ne
deskriptorov obp2ektov, a ukazatelee0 na nih, s ispolp1zovaniem
posleduyushchee0 dinamicheskoe0 rekonstruktsii samih obp2ektov.
Pri ispolp1zovanii e1toe0 vozmozhnosti v interaktivnyh 
psihoinformatsionnyh videosistemah realp1nogo vremeni dlya
uskorennoe neverbalp1noe0 kognitivnoe0 kommunikatsii,
predstavlayushchihsya avtoru v silu mnogih prichin, nekotorye
iz kotoryh stanut yasnymi v protsesse izlozheniya materiala 
statp1i, naibolee perspektivnymi, svyazp1 mezhdu ukazatelyami i
samimi obp2ektami, a takzhe protsess dinamicheskoe0 rekonstruktsii,
osushchestvlyaemoe0 v {\cyi realp1nom vremeni\/} i, kak sledstvie,
po krae0nee0 mere otchasti, na bezsoznatelp1nom urovne, chto, v 
tselom, sostavlyaet ee harakternuyu chertu, otlichayushchuyu
dinamicheskuyu rekonstruktsiyu droe1mov ot obychnoe0
re\-kons\-t\-ruk\-tsii-``rasshifrovki'', stanovyat\-sya netrivialp1nymi,
potentsialp1no vesp1ma so\-der\-zha\-telp1\-ny\-mi kak e1ksperimentalp1no,
tak i teoreticheski. Odnako, imenno znanie matematicheskogo 
fundamenta, lezhashchego v ih osnove, s odnoe0 storony, mozhet
obespechitp1 znachitelp1noe snizhenie chrezmernyh trebovanie0
k apparatnoe0 baze sistem, pozvoliv realizovatp1 ih na osnove
imeyushchihsya v nalichii i obshchedostupnyh sredstv, a s drugoe0
storony, mozhet sluzhitp1 osnovaniem dlya razrabotki udobnogo i
sravnitelp1no prostogo programmnogo obespecheniya.

Vvidu skazannogo v dannoe0 rabote my pytaemsya vyyavitp1 
matematicheskie osnovy otmechennogo vyshe podhoda k organizatsii
informatsii i ee peredachi v interaktivnyh psihoinformatsionnyh
videosistemah realp1nogo vremeni dlya uskorennoe0 neverbalp1noe0
kognitivnoe0 kommunikatsii; na e1tom puti klyuchevuyu rolp1 igrayut
droe1my (``dinamicheski rekonstruiruemye obp2ekty e1ksperimentalp1noe0
matematiki'') i interpretatsionnye figury kak ukazateli na nih.

Statp1ya organizovana sleduyushchim obrazom: pervye0 paragraf
posvyashchen iz\-lo\-zhe\-niyu neobhodimyh osnov interpretatsionnoe0
geometrii, vtoroe0 -- operatornym metodam v teorii interaktivnyh
dinamicheskih videosistem i nekommutativnoe0 nachertatelp1noe0
geometrii, tretie0 -- obshchee0 kontseptsii organizatsii
integrirovannyh interaktivnyh videokognitivnyh sistem realp1nogo
vremeni, chetvertye0 -- dinamicheski rekonstruiruemym obp2ektam
e1ksperimentalp1noe0 ma\-te\-ma\-ti\-ki i protsessam ih dinamicheskoe0
rekonstruktsii, pri e1tom obshchie ponyatiya illyustriruyut\-sya
interesnym konkretnym primerom, svyazannym s bes\-ko\-nech\-no\-mer\-noe0
geometriee0. Izlozhenie nosit, v osnovnom, e1vristicheskie0 i
kontseptualp1nye0 harakter (pervye0 i tretie0 paragrafy), hotya
nekotorye chastnye aspekty takie, kak, naprimer, material vtorogo 
i chetvertogo paragrafa, dopuskayushchie v nastoyashchie0 moment
bolee glubokuyu formalizatsiyu i detalizatsiyu, izlagayut\-sya na
matematicheskom urovne strogosti.

Otmetim takzhe, chto osnovaniem dlya otdelp1nyh polozhenie0,
izlagaemyh v ra\-bo\-te, posluzhili konkretnye resheniya,
ispolp1zuemye v razlichnyh obuchayushchih i kommunikatsionnyh
sistemah, odnako, posledovatelp1nye0 i sistematicheskie0
e1k\-s\-pe\-ri\-men\-talp1\-no-matematicheskie0 podhod k teme soderzhit\-sya
v literature vpervye.

\head\cyb\S 1. Interpretatsionnaya geometriya\endhead

\cyr Dannye0 paragraf posvyashchen izlozheniyu razlichnyh storon
interpretatsionnoe0 geometrii, imeyushchih otnoshenie k
organizatsii informatsii i ee peredachi v interaktivnyh
videosistemah kompp1yuternoe0 i telekommunikatsii. Osnovnoe 
vnimanie udelyaet\-sya interpretatsionnym figuram, kotorye v
dalp1nee0shem budut sluzhitp1 ukazatelyami na droe1my i sposobam
ih peredachi, inymi slovami, oso\-ben\-nos\-tyam interpretatsionnoe0
geometrii v mnogopolp1zovatelp1skom rezhime.

\subhead\cyb 1.1. Interpretatsionnye figury [1,2:Prilozh.A]\endsubhead
\cyr Geometriya, opisyvaemaya nizhe, svyazana s nekotorym klassom
interaktivnyh informatsionnyh sistem. Budem nazyvatp1 interaktivnuyu
informatsionnuyu sistemu kompp1yuterograficheskoe0 (ili interaktivnoe0
informatsionnoe0 videosistemoe0), esli informatsionnye0 potok
``kompp1yuter-polp1zovatelp1'' organizovan kak potok geometricheskih
gra\-fi\-ches\-kih dannyh na e1krane monitora; interaktivnaya
informatsionnaya videosistema budet nazyvatp1sya psihoinformatsionnoe0,
esli informatsiya, peredavaemaya po kanalu ``polp1zovatelp1-kompp1yuter''
nosit (polnostp1yu ili otchasti) bessoznatelp1nye0 harakter. Takim
obrazom, v interesuyushchih nas interaktivnyh sistemah upravlenie
spareno s neizvestnoe0 ili ne vpolne izvestnoe0 obratnoe0 svyazp1yu,
sistemy s takim upravleniem budem nazyvatp1 {\cyi interaktivno
upravlyaemymi}. V tselom, rassmotrenie interaktivno upravlyaemyh
(psihoinformatsionnyh) sistem, s tselp1yu kak e1ksperimentalp1nogo,
tak i teoreticheskogo vyyavleniya zaklyuchennyh v nih 
vozmozhnostee0, predstavlyayushchih interes dlya 
matematicheskih nauk samih po sebe, i ``skrytyh'' abstraktnyh
matematicheskih obp2ektov, nablyudenie i analiz kotoryh
aktualp1no ili potentsialp1no realizuemy blagodarya ukazannym
vozmozhnostyam, yavlyaet\-sya vazhnoe0 zadachee0 kak takovoe0.
Poe1tomu dalee budut opredeleny ponyatiya interpretatsionnoe0 
figury i ee simvolicheskogo cher\-te\-zha, kotorye, po-vidimomu, igrayut
klyuchevuyu rolp1 v opisanii kom\-pp1yu\-ter\-no-geo\-met\-ri\-ches\-ko\-go
predstavleniya matematicheskih dannyh v interaktivnyh informatsionnyh
sistemah. V dalp1nee0shem, odnako, pri sohranenii obshchego
e1ks\-pe\-ri\-men\-talp1\-no-matematicheskogo plana aktsenty budut smeshcheny
neskolp1ko v sto\-ro\-nu prilozhenie0 k informatike, interpretatsionnye
figury (sm.nizhe) budut ispolp1zovatp1sya kak ukazateli na droe1my, 
a interaktivnye psihoinformatsionnye videosistemy realp1nogo vremeni
kak komponenty integrirovannyh interaktivnyh videokognitivnyh sistem
dlya uskorennoe0 neverbalp1noe0 kognitivnoe0 kommunikatsii.

Matematicheskie dannye v interaktivnyh informatsionnyh 
videosistemah su\-shches\-t\-vu\-yut v vide vzaimosvyazi vnutrennego
geometricheskogo obraza (figury) v subp2ektivnom prostranstve
nablyudatelya (polp1zovatelya) i vneshnego
kom\-pp1yu\-te\-ro\-gra\-fi\-ches\-ko\-go
predstavleniya, pri e1tom vneshnee predstavlenie vklyuchaet 
vidimye e1lementy (chertezhi figur), a takzhe nevidimye
e1lementy (ana\-li\-ti\-ches\-kie vyrazheniya i algoritmy dlya
postroeniya dannyh chertezhee0). Protsess sopostavleniya
geometricheskogo obraza (figury) vo vnutrennem prostranstve
na\-blyu\-da\-te\-lya vneshnemu predstavleniyu (vidimym i nevidimym
e1lementam) budet nazyvatp1sya {\cyi translyatsiee0}, pri
e1tom vidimye0 obp2ekt mozhet bytp1 netozhdestven samoe0 figure,
v e1tom sluchae chastnye vidimye e1lementy mogut rassmatrivatp1sya 
kak moduli, chp1ya translyatsiya realizuet\-sya nezavisimo, vvidu
chego budem nazyvatp1 translyatsiyu {\cyi interpretatsiee0}, esli
translyatsiya chastnyh modulee0 realizuet\-sya v zavisimosti ot
rezulp1tata translyatsii predydushchih, i {\cyi kompilyatsiee0\/} v 
protivnom sluchae.

\definition{\cyb Opredelenie 1} \cyr Figura, poluchaemaya kak
rezulp1tat interpretatsii, nazyvaet\-sya {\cyi interpretatsionnoe0
figuroe0}.\enddefinition

\cyr Podcherknem, chto interpretatsionnaya figura mozhet ne imetp1
obychnogo formalp1nogo opredeleniya; a imenno, tolp1ko esli protsess
interpretatsii dopuskaet e1kvivalentnye0 protsess kompilyatsii
opredelenie figury svoditsya k opredeleniyu ee chertezhee0, chto v
obshchem sluchae ne imeet mesto. tem samym, chertezh 
interpretatsionnoe0 figury opredelyaet vsego lishp1 dinamicheskuyu
``tehnologiyu zritelp1nogo vospriyatiya'', a ne ee ``obraz'';
podobnye chertezhi budut nazyvatp1sya {\cyi simvolicheskimi}.

Kompp1yuterno-geometricheskoe opisanie matematicheskih dannyh v
in\-te\-rak\-tiv\-nyh informatsionnyh sistemah tesno svyazano s
kontseptsiee0 anomalp1nyh virtualp1nyh realp1nostee0.

\subhead\cyb 1.2. Intentsionalp1nye anomalp1nye virtualp1nye realp1nosti
[1,2:Pri\-lozh.A]\endsubhead

\definition{\cyb Opredelenie 2 {\cyr (sr.[1,2:Prilozh.A])}} \cyr
{\cyb (A).} {\cyi Anamalp1noe0 virtualp1noe0 realp1nostp1yu\/}
({\cyi AVR\/}) {\cyi v uzkom smysle\/} nazyvaet\-sya nekotoraya
opredelennaya sistema pravil ne\-stan\-dart\-noe0 nachertatelp1noe0
geometrii, prisposoblennoe0 k realizatsii na vi\-deo\-kom\-pp1yu\-te\-re
(ili mulp1tisensornoe0 sisteme ``virtualp1noe0 realp1nosti'' [3-6]).
{\cyi Anomalp1naya virtualp1naya realp1nostp1 v shirokom smysle\/}
vklyuchaet v sebya tak\-zhe izobrazhenie v kiberprostranstve,
vypolnennoe soglasno ukazannoe0 sisteme pravil. V dalp1nee0shem
termin budet ispolp1zovatp1sya v uzkom smysle.
{\cyb (B).} {\cyi Na\-tu\-ra\-li\-za\-tsi\-ee0\/} nazyvaet\-sya protsess
sopostavleniya anomalp1noe0 virtualp1noe0 re\-alp1\-nos\-ti nekotoroe0
abstraktnoe0 geometrii ili fizicheskoe0 modeli. Budem go\-vo\-ritp1,
chto anomalp1naya virtualp1naya realp1nostp1 {\cyi naturalizuet\/}
abstraktnuyu mo\-delp1, a modelp1 {\cyi transtsendiruet\/}
naturalizuyushchuyu ee anomalp1nuyu virtualp1nuyu realp1nostp1.
{\cyb (V).} {\cyi Vizualizatsiee0\/} nazyvaet\-sya protsess
sopostavleniya nekotorogo izobrazheniya ili vizualp1noe0 dinamiki
v nekotoroe0 anomalp1noe0 virtualp1noe0 realp1nosti (realizovannoe0
apparatno i programmno kompp1yuterograficheskim interfee0som
konkretnoe0 videosistemy) obp2ektam abstraktnoe0 geometrii ili
protsessam v fizicheskoe0 modeli.
{\cyb (G).} Anomalp1naya virtualp1naya realp1nostp1, izobrazheniya 
v kotoroe0 zavisyat ot nablyudatelya, nazyvaet\-sya {\cyi 
intentsionalp1noe0 anomalp1noe0 virtualp1noe0 realp1nostp1yu\/}
({\cyi IAVR\/}). Obobshchennye zakony perspektivy v IAVR vklyuchayut
uravneniya interaktivnoe0 dinamiki nablyudaemyh obp2ektov naryadu
so standartnymi (geometricheskimi) zakonami perspektivy. Protsess
nablyudeniya v IAVR sostoit iz fizicheskogo protsessa nablyudeniya 
i vir\-tu\-alp1\-no\-go protsessa intendirovaniya, kotorye0 upravlyaet
e1volyutsiee0 izobrazhenie0 soglasno dinamicheskim zakonam
perspektivy.
\enddefinition

\cyr V intentsionalp1noe0 anomalp1noe0 virtualp1noe0 realp1nosti
(IAVR), realizovannoe0 (aparatno i programmno) {\cyi
kompp1yuterograficheskim interfee0som in\-te\-rak\-tiv\-noe0 videosistemy},
kotorye0 v svoyu ocheredp1 ukazannaya IAVR {\cyi geo\-met\-ri\-ches\-ki
modeliruet\/} (na urovne nachertatelp1noe0 geometrii, v to vremya
kak modelp1, transtsendiruyushchaya dannuyu IAVR, osushchestvlyaet 
e1to uzhe na urovne abst\-rakt\-noe0 geometrii), nablyudaemye obp2ekty
predstayut kak by svyazannymi s nablyudatelem, kotorye0
opredelennym obrazom vozdee0stvuya na nih, fiksiruet ih nab\-lyu\-da\-e\-moe
sostoyanie, tak chto obp2ekt myslit\-sya tolp1ko kak 
potentsialp1nostp1 so\-sto\-ya\-niya iz zadannogo spektra, no samo
sostoyanie zavisit i ot nablyudatelya. Simvolicheskie chertezhi
interpretatsionnyh figur mogut rassmatrivatp1sya kak sostoyaniya
nekotoroe0 IAVR.

Otmetim, chto matematicheskaya teoriya anomalp1nyh virtualp1nyh
realp1nostee0 (AVR), v tom chisle osnovnye protsedury naturalizatsii
i transtsendirovaniya, svyazyvayushchie AVR s abstraktnoe0 geometriee0,
predstavlyaet soboe0 osobye0 raz\-del {\cyi sovremennoe0 neklassicheskoe0
nachertatelp1noe0 (kompp1yuternoe0) geometrii}.

\definition{\cyb Opredelenie 2D} \cyr Mnozhestvo nepreryvno
raspredelennyh vizualp1nyh harakteristik izobrazheniya v anomalp1noe0
virtualp1noe0 realp1nosti nazyvaet\-sya {\cyi ano\-malp1\-nym tsvetovym
prostranstvom\/}; e1lementy anomalp1nogo tsvetovogo prost\-ran\-st\-va
netsvetovoe0 prirody nazyvayut\-sya {\cyi obertsevetami\/}, a 
velichiny, trans\-tsen\-di\-ru\-yu\-shchie ih v abstraktnoe0 modeli,
nazyvayut\-sya {\cyi ``skrytymi svetami''}. 
{\cyi Tsve\-to\-pers\-pek\-tiv\-noe0 sistemoe0\/} nazyvaet\-sya zadannaya
sovokupnostp1 obobshchennyh zakonov perspektivy v zadannom 
anomalp1nom tsvetovom prostranstve; dve AVR nazyvayut\-sya 
e1kvivalentnymi, esli ih tsvetoperspektivnye sistemy sovpadayut;
AVR, e1k\-vi\-va\-lent\-naya realizovannoe0 na videokompp1yutere (no
ne realizovannaya sama) nazyvaet\-sya {\cyi marginalp1noe0}.
\enddefinition

\subhead\cyb 1.3. Nealeksandrie0skaya interpretatsionnaya
geometriya [7,2:Prilozh.A]\endsubhead \cyr Ot\-me\-tim, chto
bolp1shinstvo klassicheskih geometrie0 ispolp1zuet nekotorye0
postulat, kotorye0 my budem nazyvatp1 aleksandrie0skim, no ne
vklyuchaet yavno ego v svoyu aksiomatiku.

\definition{\cyb Aleksandrie0skie0 postulat} \cyi Lyuboe
utverzhdenie, spravedlivoe dlya nekotoroe0 geometricheskoe0
konfiguratsii, ostaet\-sya v sile, esli e1ta konfiguratsiya
ras\-smat\-ri\-va\-et\-sya kak chastp1 nekotoroe0 bolp1shee0 konfiguratsii.
\enddefinition

\cyr Takim obrazom, aleksandrie0skie0 postulat oznachaet, chto
dobavlenie kakih by to ni bylo dopolnitelp1nyh obp2ektov k
dannoe0 geometricheskoe0 konfiguratsii ne okazyvaet vliyaniya
na e1tu konfiguratsiyu. Aleksandrie0skie0 postulat 
ispolp1zuet\-sya, naprimer, v konstruktivnyh dokazatelp1stvah
geometricheskih teorem, osushchestvlyaemyh putem yavnogo 
poshagovogo postroeniya obp2ekta, su\-shches\-t\-vo\-va\-nie kotorogo
sostavlyaet utverzhdenie teoremy. V sluchae narusheniya 
aleksandrie0skogo postulata v protsesse konstruktivnogo
dokazatelp1stva na kazhdom shage mogut ispolp1zovatp1sya
tolp1ko konfiguratsii, sovmestimye v nekotorom smysle s
predydushchimi, t.e. ne menyayushchimi te ih svoe0stva, kotorye
sostavlyayut is\-hodnye dannye teoremy.

V kachestve primera nealeksandrie0skoe0 geometrii mozhno privesti
e1e0nsh\-tee0\-no\-vu geometriyu [7,2:Prilozh.A]. Kinematika i protsess
rasseyaniya figur mogut bytp1 prodemonstrirovany na drugom vazhnom
primere nealeksandrie0skoe0 geo\-met\-rii -- geometrii solitonov
[7,2:Prilozh.A]. Vse primery solitonnyh geo\-met\-rie0 podtverzhdayut
predpolozhenie, chto narushenie aleksandrie0skogo postulata svyazano
s vzaimodee0stviem geometricheskih figur, i, v chastnosti, podobnoe
vzaimodee0stvie mozhet opredelyatp1sya nelinee0nym harakterom
e1volyutsii.

Rassmotrim protsess rasseyaniya figur v interpretatsionnoe0
geometrii. Kak otmechalosp1 vyshe, interpretatsionnaya figura
opisyvaet\-sya paroe0 $(\Phi^{\tni},\Phi^{\txe})$, gde $\Phi^{\tni}$ 
-- vnutrennie0 obraz v subp2ektivnom prostranstve nablyudatelya, a
$\Phi^{txe}$ -- ego chertezh; $\Phi^{\tni}$ -- rezulp1tat
interpretatsii $\Phi^{\txe}$. Estestvenno predpolozhitp1, chto
$\Phi^{\tni}$ zavisit ot $\Phi^{\txe}$ funktsionalp1no:
$\Phi^{\tni}_t=\Phi^{\tni}\left[\Phi^{\txe}_{\tau\le t}\right]$ i, 
kak pravilo, nelinee0no; bolee togo, esli $\Phi^{\txe}$ obladaet
asimptoticheskoe0 svobodoe0, to $\Phi^{\tni}$ takzhe obladaet eyu.
Kak sledstvie, dazhe pri linee0noe0 dinamike $\Phi^{\txe}$ imeet
mesto netrivialp1noe rasseyanie interpretatsionnyh figur, takim
obrazom, interpretatsionnye figury yavlyayut\-sya
nealeksandrie0skimi.

Takim obrazom, interpretatsionnye geometrii realizuyut nekotorye0 
klass nealeksandrie0skih geometrie0. Konstruktivnoe 
dokazatelp1stvo v in\-ter\-pre\-ta\-tsi\-on\-noe0 geometrii, tem samym,
sostoit v yavnom postroenii obp2ekta, chp1e sushchestvovanie
utverzhdaet teorema, kak rezulp1tata poshagovogo protsessa
postroeniya pro\-me\-zhu\-toch\-nyh konfiguratsie0 interpretatsionnyh
figur, a korrektnostp1 dokazatelp1stva zaklyuchaet\-sya v tom,
chto na kazhdom shagu svoe0stva nachalp1noe0 konfiguratsii,
sostavlyayushchie is\-hodnye dannye teoremy, osta\-yut\-sya
neizmennymi v pro\-tses\-se interpretatsii. Esli poslednee
proveryaet\-sya e1ksperimentalp1no dlya kazh\-do\-go
konkretnogo protsessa, to dokazatelp1stvo nosit
e1ksperimentalp1no-ma\-te\-ma\-ti\-ches\-kie0 harakter, buduchi osnovannym
na nepolnoe0 induktsii; otmetim, chto {s}hod\-nye protsedury
ispolp1zuyut\-sya vo mnogih samoobuchayushchihsya sistemah
is\-kus\-st\-ven\-no\-go intellekta, v kotoryh realizuyut\-sya (inogda
statisticheski) voz\-mozh\-nye kombinatsii shagov s posleduyushchee0
proverkoe0 ih korrektnosti (metod slu\-chae0\-no\-go poiska reshenie0).

\subhead\cyb 1.4. Informatsionnye aspekty interpretatsionnyh
geometrie0: in\-ter\-pre\-ta\-tsi\-on\-nye figury v mnogopolp1zovatelp1skom
rezhime i IAVR-telesteziya\endsubhead \cyr V dannom punkte
soderzhit\-sya kratkoe izlozhenie bez detalizatsii obshchih
re\-zulp1\-ta\-tov vtoroe0 chasti raboty [7], ne voshedshih v prilozhenie
k rabote [2] i ka\-sa\-yu\-shchih\-sya informatsionnyh aspektov
interpretatsionnyh geometrie0. Predposylkoe0 dlya takogo 
rassmotreniya sluzhit otmechavsheesya v [7] obstoyatelp1stvo, chto
informatika mozhet rassmatrivatp1sya kak tochka zreniya na
matematicheskie obp2ekty, dopolnitelp1naya k geometricheskoe0
(v svete fundamentalp1noe0 oppozitsii ``lo\-gi\-ches\-ko\-go'' i
``e1e0deticheskogo''). V silu e1togo polezno pereformulirovatp1
osnovnye geometricheskie opredeleniya v informatsionnyh
terminah. Tak, opisannye vo vtoroe0 chasti raboty [7]
ponyatiya AVR-fotodosii i ee formalp1noe0 grammatiki
predstavlyayut estestvennuyu parallelp1 k ponyatiyam anomalp1noe0
vir\-tu\-alp1\-noe0 realp1nosti i transtsendiruyushchee0 abstraktnoe0
modeli. Otmetim, odnako, chto ukazannye ponyatiya v silu
metodologicheskih soobrazhenie0 rassmatrivayut\-sya poka lishp1
chisto formalp1no i kak realizuyushchie ``nizhnie0'' ili ``vneshnie0'
vizualp1nye0 urovenp1 ``mnogosloe0nogo' videokognitivnogo 
smyslovogo potoka. Ih bolee glubokomu (no daleko ne polnomu)
analizu budet posvyashchen chetvertye0 paragraf raboty.

\definition{\cyb Opredelenie 3A} \cyr Peredacha informatsii cherez
anomalp1nuyu virtualp1nuyu re\-alp1\-nostp1 s pomoshchp1yu ``skrytyh
svetov'' nazyvaet\-sya {\cyi AVR-fotodosiee0\/}; sistema
algebraicheskih struktur iznachalp1noe0 abstraktnoe0 modeli,
harakterizuyushchih pri naturalizatsii protsess AVR-fotodosii,
nazyvaet\-sya {\cyi formalp1noe0 grammatikoe0\/} AVR-fotodosii.
\enddefinition

\cyr Otmetim, chto ponyatiya AVR-fotodosii i ee formalp1noe0
tesno svyazany s po\-nya\-ti\-em anomalp1nogo tsvetovogo prostranstva,
poskolp1ku imenno ispolp1zovanie podobnyh prostranstv pozvolyaet
peredavatp1 raznoobraznuyu informatsiyu v raz\-lich\-nyh formah, i
kak sledstvie, izuchenie problem peredachi informatsii cherez 
AVR, harakter kotoroe0 zavisit ot struktury tsvetovogo
prostranstve, yavlyaet\-sya vazhnoe0 matematicheskoe0 problemoe0
(sr.[8]). Struktura AVR-fotodosii opre\-de\-lya\-et\-sya ee formalp1noe0
grammatikoe0. Kak budet pokazano v sleduyushchem pa\-rag\-ra\-fe
formalp1naya grammatika interpretatsionnyh geometrie0 mozhet
nositp1 kvantovye0 harakter i, takim obrazom, sootvet\-stvuyushchaya
AVR-fotodosiya podchinyaet\-sya kvantovoe0 logike [9]; e1tot fakt
zasluzhivaet samogo pristalp1nogo vnimaniya kak teoreticheski
poluchennye0 smeshannye0 pertseptivno-kognitivnye0 analog vesp1\-ma
interesnyh i t{shch}atelp1no izuchaemyh nauchnymi kollektivami
matematikov, fi\-zi\-kov, psihologov i nee0rofiziologov (kak
teoreticheski, tak i e1ks\-pe\-ri\-men\-talp1\-no) chisto kognitivnyh
protsessov, udovletvoryayushchih kvantovoe0 logike. Otmetim, chto
formalp1naya grammatika AVR-fotodosii v nekotoryh konkretnyh modelyah
interpretatsionnyh geometrie0 obsuzhdalasp1 vo vtoroe0 chasti raboty
[7] s tsitirovaniem mnogochislennoe0 literatury.

Neobhodimo otmetitp1, chto IAVR, kak pravilo, polisemantichna;
e1to ozna\-cha\-et, chto obp2em i struktura informatsii, poluchaemoe0
pri AVR-fotodosii cherez nee, zavisit ot nablyudatelya; takim
obrazom, estestvennaya problema sostoit v opi\-sa\-nii informatiki
interaktivnyh psihoinformatsionnyh videosistem, so\-der\-zha\-shchih
bolee odnogo nablyudatelya, v chastnosti, korrelirovannostp1
razlichnyh nablyudenie0. Podobnye sistemy mogut rassmatrivatp1sya 
kak realizuyushchie in\-te\-rak\-tiv\-nuyu {\rm MISD
(Multiple\_Instruction--Single\_Data)} arhitekturu s 
pa\-ral\-lelp1\-ny\-mi interpretatsionnymi protsessami dlya razlichnyh
nablyudatelee0 (e1tot fakt sleduet rassmatrivatp1 v kontekste
zamechaniya o kvantovo-logicheskom haraktere AVR-fotodosii); na
e1tom puti my stalkivaemsya s yavleniem, spetsifichnym dlya
podobnyh sistem, no, vozmozhno, imeyushchim i bolee obshchee
znachenie: a imenno, protsessy nablyudeniya razlichnymi
nablyudatelyami porozhdayut informatsionnye0 obmen mezhdu nimi.

\definition{\cyb Opredelenie 3B} \cyr AVR-fotodosiya cherez 
IAVR ot odnogo nablyudatelya k drugomu nazyvaet\-sya {\cyi 
IAVR-telesteziee0\/}; esli v protsesse IAVR-telestezii 
AVR-fotodosii ot razlichnyh nablyudatelee0 ne 
udovletvoryayut printsipu su\-per\-po\-zi\-tsii, budem govoritp1, chto
imeet mesto {\cyi kollektivnye0 e1ffekt\/} v IAVR-telestezii.
\enddefinition

\cyr Otmetim, chto (1) protsess IAVR-telestezii nosit
dvustoronnie0 harakter; nablyudateli, uchastvuyushchie v
IAVR-telesteticheskoe0 kommunikatsii, vy\-stu\-pa\-yut odnovremenno
i kak induktory (posylayushchie informatsiyu), i kak
retsepienty (prinimayushchie ee), bolee togo, obp2em i struktura
poluchaemoe0 in\-for\-ma\-tsii zavisit ot retsepienta tak zhe kak i ot
induktora; (2) kollektivnye0 e1ffekt v IAVR-telestezii oznachaet,
chto induktory v IAVR ne vosprinimayut\-sya kak nezavisimye --
peredavaemaya informatsiya ne yavlyaet\-sya summoe0 informatsie0,
posylaemyh otdelp1nymi nablyudatelyami, tak kak partsialp1nye
informatsionnye potoki ot kazhdogo induktora uchastvuyut v
obmennom vzaimodee0stvii, formiruya spetsificheskuyu informatsiyu,
vosprinimaemuyu retsepientom. V e1tom aspekte dolzhno bytp1
osobo otmecheno otnoshenie prois\-hozhdeniya IAVR-telestezii k
to\-mu faktu, chto interaktivnye psihoinformatsionnye videosistemy
realizuyut interaktivnuyu {\rm MISD} arhitekturu.

\definition{\cyb Opredelenie 3V} \cyr Nablyudatelp1 v marginalp1noe0
AVR, kotoromu ne so\-ot\-vet\-st\-vu\-et nikakoe0 nablyudatelp1 v AVR,
realizovannoe0 na videokompp1yutere, na\-zy\-va\-et\-sya {\cyi
virtualp1nym\/}; a virtualp1nye0 nablyudatelp1, chee0 protsess
nablyudeniya zavisit ot neskolp1kih realp1nyh nablyudatelee0,
nazyvaet\-sya {\cyi kollektivnym virtualp1nym nablyudatelem}.
\enddefinition

\cyr Prisut\-stvie virtualp1nogo nablyudatelya oznachaet, chto
chastp1 prinimaemoe0 informatsii interpretiruet\-sya kak
informatsiya, posylaemaya ukazannym realp1no ne 
sushchestvuyushchim nablyudatelem. Prisut\-stvie kollektivnogo
virtualp1nogo nab\-lyu\-da\-te\-lya ne obyazatelp1no, no obychno dlya
interaktivnyh videosistem v mnogopolp1zovatelp1skom rezhime;
e1tot fakt takzhe sleduet rassmatrivatp1 v kontekste zamechaniya 
o tom, chto ukazannye sistemy realizuyut in\-te\-rak\-tiv\-nuyu
virtualp1nuyu {\rm MISD} arhitekturu s pa\-ral\-lelp1\-ny\-mi
interpretatsionnymi protsessami dlya razlichnyh nablyudatelee0.

Vo vtoroe0 chasti raboty [7] izlozhennye obshchie ponyatiya
proillyustrirovany na konkretnoe0 modeli interpretatsionnoe0
geometrii, v chastnosti, dan primer kollektivnogo virtualp1nogo
nablyudatelya, proyasnyayushchie0 ego smysl. Tak ot\-me\-cha\-et\-sya,
chto (1) tolp1ko chastp1 prinimaemoe0 informatsii interpretiruet\-sya
kak informatsiya, posylaemaya kollektivnym virtualp1nym nablyudatelem
(t.o. ego na\-li\-chie ne narushaet prisut\-stviya realp1nyh nablyudatelee0),
(2) protsess in\-ten\-di\-ro\-va\-niya kollektivnogo virtualp1nogo
nablyudatelya vsetselo opredelyaet\-sya vzai\-mo\-dee0\-st\-viem realp1nyh
nablyudatelee0 (t.e. kollektivnye0 virtualp1nye0 nablyudatelp1 predstavlyaet
soboe0 spetsificheskoe obp2edinennoe sostoyanie realp1nyh
na\-b\-lyu\-da\-te\-lee0 v interaktivnoe0 psihoinformatsionnoe0
videosisteme v mno\-go\-polp1\-zo\-va\-telp1\-s\-kom rezhime), (3)
kollektivnye0 virtualp1nye0 nablyudatelp1 uchast\-vu\-et v informatsionnom
obmene s realp1nymi nablyudatelyami, buduchi interpretiruemym (po krae0nee0
mere, formalp1no) kak nezavisimye0 nablyudatelp1. V svyazi s e1tim
predstavlyaet\-sya vesp1ma vazhnym pochti ne issledovannye0 vopros 
o vzaimodee0stvii individualp1nyh realp1nyh nablyudatelee0 s 
kollektivnym virtualp1nym nablyudatelem, a takzhe o razlozhenii
poslednego na nekorrelirovannye sostavlyayushchie ({\cyi
kvazisubp2ekty}) i ih vzaimodee0stviyah mezhdu soboe0.

V zaklyuchenie sformuliruem predlozhenie, dokazyvaemoe v kontse
vtoroe0 chas\-ti raboty [7] putem yavnogo postroeniya.

\proclaim{\cyb Predlozhenie} {\cyi Sushchestvuyut modeli
interpretatsionnyh geometrie0, v kotoryh ime\-yut\-sya
interpretatsionnye fugury, nablyudaemye tolp1ko v
mno\-go\-polp1\-zo\-va\-telp1\-s\-kom rezhime}.
\endproclaim

\cyr Predstavlyaet interes vyyavlenie znacheniya mehanizma
nablyudeniya podobnyh figur v ramkah obshchee0 
matematicheskoe0 teorii igr [10,11].

K sozhaleniyu, nesmotrya na svoi dostoinstva opisannye0 vyshe
mehanizm IAVR-telestezii, po-vidimomu, ne pozvolyaet
neposredstvenno osushchestvlyatp1 peredachu kognitivnoe0
informatsii. Opisaniyu popytki tem ne menee ispolp1zovatp1
IAVR-telesteziyu dlya realizatsii uskorennoe0 neverbalp1noe0
kognitivnoe0 kom\-mu\-ni\-ka\-tsii v ramkah integrirovannyh {\cyi
videokognitivnyh\/} interaktivnyh sistem po\-svya\-shche\-ny
tretie0 i chetvertye0 paragraf dannoe0 statp1i.

\head\cyb\S 2. Operatornye (kvantovo-polevye i stohasticheskie)
metody v teorii interaktivnyh dinamicheskih videosistem i
nekommutativnaya nachertatelp1naya geometriya\endhead

\cyr Danye0 paragraf posvyashchen formalizatsii i detalizatsii
izlozhennoe0 v predydushchem paragrafe intuitivno prozrachnoe0
geometricheskoe0 kartiny i razrabotke sootvet\-stvuyushchego
algebraicheskogo i analiticheskogo apparata. V rezulp1tate
usushchestvleniya e1togo my budem obladatp1 vozmozhnostp1yu
bolee strogogo ma\-te\-ma\-ti\-ches\-ko\-go opisaniya konkretnyh modelee0
kak s tselp1yu dalp1nee0shego proyasneniya obshchih 
geometricheskih aspektov, tak i konkretnoe0 
apparatno-programmnoe0 raz\-ra\-bot\-ki konkretnyh interaktivnyh
videosistem realp1nogo vremeni.

\subhead\cyb 2.1. Obshchie operatornye aspekty interaktivnyh
dinamicheskih videosistem i nekommutativnaya geometriya\endsubhead
\cyr Sushchestvuet neskolp1ko shiroko upotreblyaemyh obshchih 
sposobov zadaniya e1volyutsii izobrazhenie0 v interaktivnyh
dinamicheskih videosistemah realp1nogo vremeni. Privedem
nekotorye iz nih:
\roster
\item {\cyi Formuly E1e0lera\/} [12]:
$$\dot\Phi(t)=A(t,u,\dot u,\xi)\Phi(t),$$
gde $u=u(t)\in\Bbb C\simeq\Bbb R^2$ -- tekushchee polozhenie tochki
vzora (takim obrazom, e1kran rassmatrivaet\-sya kak chastp1
kompleksnoe0 ploskosti $\Bbb C\simeq\Bbb R^2$), $\dot u=\dot u(t)$ -- 
otnositelp1naya skorostp1 ee smeshcheniya, $\xi=\xi(t)$ -- 
dopolnitelp1nye dinamicheskie parametry interaktivnogo
upravleniya, $\Phi=\Phi(t)$ -- sovokupnostp1 nepreryvno 
raspredelennyh kiralp1nyh (t.e. 
golomorfno-antigolomorfno rasshcheplennyh, sm.[1]) vizualp1nyh
harakteristik izobrazheniya (tsvetov i obertsvetov),
$A=A(t,u,\dot u,\xi)$ -- linee0nye0 operator. Linee0nye0 operator $A$
kak funktsiya $u$ i $\dot u$ nazyvaet\-sya {\cyi uglovym
operatornym polem}, pole $A$ (voobshche govorya, neogranichennoe)
golomorfno zavisit ot $u$ i $\dot u$ i (slabo) nepreryvno po $\xi$ 
v podhodyashchee0 (ne obyazatelp1no met\-ri\-zu\-e\-moe0) obshchee0 topologii
na prostranstve parametrov interaktivnogo uprav\-le\-niya (naprimer,
biopotentsialov E1E1G i E1RG, dinamicheskih pa\-ra\-met\-rov
respiratornogo ritma i t.p., a takzhe funktsionalp1nyh
komp\-lek\-sov perechislennyh velichin). Dinamika uglovogo
operatornogo polya po peremennoe0 $t$ mozhet v svoyu ocheredp1
udovletvoryatp1 nekotoromu dif\-fe\-ren\-tsi\-alp1\-no\-mu uravneniyu
(naprimer, uravneniyu E1e0lera-Arnolp1da [13]).
\item {\cyi Formuly E1e0lera-Be\-lav\-ki\-na-Kolokolp1tsova\/} [2]:
$$d\Phi(t,[\omega])=A(t,u,\dot u,\xi)\Phi(t)dt+\sum_{\alpha}B_\alpha(t,u,\dot
u,\xi)\Phi(t,[\omega])d\omega^{(\alpha)},$$
gde $d\omega^{(\alpha)}$ -- nabor stohasticheskih differentsialov.
Pri e1tom, inogda na prak\-ti\-ke polya $A(u,\dot u)$ i $B_\alpha(u,\dot u)$
mogut vklyuchatp1 (slabye) nelinee0nosti. Di\-na\-mi\-ka polee0 po
peremennoe0 $t$ mozhet v svoyu ocheredp1 kak i v
de\-ter\-mi\-ni\-ro\-van\-nom sluchae udovletvoryatp1 nekotoromu
differentsialp1nomu uravneniyu tipa uravnenie0 E1e0lera-Arnolp1da
[13].
\item {\cyi Modeli s dinamicheskim interaktivnym e1kranirovaniem\/}
[1]: V e1tih modelyah sovokupnostp1 tsvetov i obertsvetov 
$\Psi=\Psi(t)$ predstavlyaet\-sya v vide
$$\Psi(t)=J(t,u,\dot u,\xi)\Phi(t),$$
gde $\Phi\!=\!\Phi(t)$ udovletvoryaet formulam E1e0lera ili formulam
E1e0lera-Be\-lav\-ki\-na-Kolokolp1tsova, a $J\!=\!J(t,u,\dot u,\xi)$ -
linee0nye0 operator (kak pra\-vi\-lo, proektor s netrivialp1nym yadrom),
kotorye0 kak (golomorfnaya) funktsiya ot $u$ i $\dot u$ nazyvaet\-sya {\cyi
e1kraniruyushchim operatornym polem}, e1k\-ra\-ni\-ru\-yu\-shchee
operatornoe pole, takzhe kak uglovoe i vse ostalp1nye polya, vhodyashchie v
uravneniya e1volyutsii izobrazheniya, (slabo) nepreryvno po $\xi$
v pod\-ho\-dya\-shchee0 (ne obyazatelp1no metrizuemoe0 i, vozmozhno,
svoee0 dlya kazhdogo ope\-ra\-tor\-no\-go polya) obshchee0 topologii
na prostranstve parametrov interaktivnogo upravleniya.
\item {\cyi Modeli s pamyatp1yu\/} [1]: V e1tom sluchae dinamika
tsvetov i obertsvetov za\-vi\-sit ot predystorii (naprimer,
yavlyaet\-sya integrodifferentsialp1noe0 po vremeni).
\endroster

\cyr Nekotorye konkretnye realizatsii dinamik, opisannyh vyshe,
dostatochno ho\-ro\-sho issledovany e1ksperimentalp1no (naprimer,
t.n. sistemy s chastichnym uvle\-che\-ni\-em i maskirovaniem -- sm.[1]).

Pri analize perechislennyh modelee0 (na ustoe0chivostp1 
izobrazheniya i t.p.) vazh\-nuyu rolp1 igrayut operatornye
(kvantovo-polevye i stohasticheskie) metody (sm.na\-pri\-mer
[1,2,7,12,13]). Ispolp1zovanie kvantovo-polevyh metodov, kak
pra\-vi\-lo, osnovyvaet\-sya na sleduyushchem dopushchenii, nazyvaemom
{\cyi gipotezoe0 ope\-ra\-tor\-noe0 algebry}, a imenno, chto
koe1ffitsienty razlozhenie0 operatornyh polee0, vho\-dya\-shchih
v dinamicheskie uravneniya, po vsem peremennym krome $u$,
obrazuyut zamknutuyu operatornuyu algebru kvantovoe0 teorii polya
buduchi operatornymi polyami po $u$. Takim obrazom, predpolagaet\-sya,
chto nekotorye velichiny, polnostp1yu ha\-rak\-te\-ri\-zu\-yu\-shchie
e1volyutsiyu sistemy, obrazuyut nekotorye0 algebraicheskie0 obp2\-ekt,
op\-re\-de\-le\-nie kotorogo privodit\-sya nizhe. Otmetim, chto v
nekotoryh chastnyh klas\-sah modelee0 gipotezu operatornoe0 algebry
udaet\-sya formalp1no dokazatp1.

\definition{\cyb Opredelenie 4A} \cyr {\cyi KTP--operatornoe0 algebroe0\/} 
({\cyi operatornoe0 algebroe0 kvantovoe0 teorii polya\/})
nazyvaet\-sya linee0noe prostranstvo $H$, snabzhennoe zavisyashchee0
ot pa\-ra\-met\-ra $\vec x$, probegayushchego $\Bbb R^n$ ili $\Bbb C^n$,
operatsiee0 $m_{\vec x}(\cdot,\cdot)$, dlya kotoroe0 vy\-pol\-nya\-et\-sya
sleduyushchee tozhdestvo:
$m_{\vec x}(\cdot,m_{\vec y}(\cdot,\cdot))\!=\!m_{\vec y}(m_{\vec
x\!-\!\vec y}(\cdot,\cdot),\cdot)$.
Inymi slovami, KTP--operatornaya algebra yavlyaet\-sya paroe0
$(H,t^k_{ij}(\vec x))$, gde $H$ -- linee0noe prostranstvo, a
$t^k_{ij}(\vec y)$ -- $H$--znachnoe tenzornoe pole na $\Bbb R^n$ 
ili $\Bbb C^n$ takoe, chto
$t^l_{im}(\vec x)t^m_{jk}(\vec y)\!=\!t^m_{ij}(\vec y\!-\!\vec
y)t^l_{mk}(\vec y)$. Pole $t_{ij}^k(\vec x)$ realizuet operatsiyu
$m_{\vec x}(\cdot,\cdot)$ sleduyushchim obrazom:
$m_{\vec x}(e_i,e_j)=t_{ij}^k(\vec x)e_k$, gde $\{e_k\}$ --
proizvolp1nye0 bazis v $H$.
\enddefinition

\cyr Vvedem operatory $l_{\vec x}(e_i)e_j=t^k_{ij}(\vec x)e_k$ 
(ope\-ra\-to\-ry umnozheniya sleva v KTP--ope\-ra\-tor\-noe0 algebre),
togda budut imetp1 mesto sleduyushchie tozhdestva:
$l_{\vec x}(e_i)l_{\vec y}(e_j)\mathbreak=t^k_{ij}(\vec
x-\vec y)l_{\vec y}(e_k)$ ({\cyi operatornoe razlozhenie\/}) i $l_{\vec
x}(e_i)l_{\vec y}(e_j)=l_{\vec y}(l_{\vec x-\vec y}(e_i)e_j)$ ({\cyi
tozh\-dest\-vo dualp1nosti\/}).

V literature po metematicheskoe0 fizike, kak pravilo, ispolp1zuet\-sya
oboznachenie $\varphi(\vec x)$ dlya operatorov $l_{\vec x}(\varphi)$
($\varphi\in H$). Velichiny $\varphi(\vec x)$ nazyvayut\-sya
ope\-ra\-tor\-ny\-mi polyami. V terminah operatornyh polee0
operatornye razlozheniya prinimayut vid
$$\varphi_1(\vec x)\varphi_2(\vec y)=F_\alpha(\vec x-\vec
y)\psi_{\alpha}(\vec y),$$
chto oznachaet razlozhimostp1 proizvedeniya operatornyh polee0 po
operatornym polyam KTP-operatornoe0 algebry. Esli nabor
konkretnyh operatornyh polee0 zamknut otnositelp1no podobnyh
razlozhenie0, to im mozhet bytp1 sopostavlena abstraktnaya
KTP-operatornaya algebra, e1lementy kotoroe0 oni predstavlyayut.

Esli $dt_{ij}^k\equiv 0$, to KTP-operatornaya algebra yavlyaet\-sya
obychnoe0 assotsiativnoe0 algebroe0. E1lement $\varphi$ 
KTP-operatornoe0 algebry nazyvaet\-sya levym delitelem nu\-lya,
esli $l_{\vec x}(\varphi)\equiv 0$; edinitsee0 v
KTP-operatornoe0 algebre $H$ nazyvaet\-sya e1le\-ment $\boldkey 1$
takoe0, chto $l_{\vec x}(\boldkey 1)\equiv\id$; v KTP-operatornoe0
algebre s edinitsee0 i bez levyh delitelee0 nulya
$\left.l_{\vec x}(\varphi)\right|_{\vec x=0}=\varphi$; 
esli $V$ -- linee0noe prostranstvo levyh delitelee0 nulya v
KTP-operatornoe0 algebre $H$, to
$(\forall\varphi\in H)l_{\vec x}(\varphi)V\subseteq V$ i 
$H/V$ -- KTP-operatornaya algebra bez levyh delitelee0 nulya.
V KTP-operatornoe0 algebra s edinitsee0 $\boldkey 1$ opredelen
operator $\bold L$ ({\cyi operator infinitezimalp1nyh 
translyatsie0\/}): $\vec\bold L\varphi=\left.\frac{d}{d\vec
x}(l_{\vec x}(\varphi)\boldkey 1)\right|_{\vec x=0}$, operator
infinitezimalp1nyh translyatsie0 $\vec\bold L$ yavlyaet\-sya
differentsirovaniem KTP-operatornoe0 algebry $H$ bez levyh
delitelee0 nulya, t.e. dlya lyubogo $\varphi$ iz $H$ imeet mesto
ravenstvo $[\vec\bold L,l_{\vec x}(\varphi)]=l_{\vec x}(\vec\bold
L\varphi)$; kak sledstvie $l_{\vec x}(\varphi)\boldkey 1=\exp(\vec
x\cdot\vec\bold L)\varphi$. Esli $H$ -- proizvolp1naya KTP-operatornaya
algebra i $\vec\bold L$ -- nekotoroe ee differentsirovanie, to v
linee0nom prostranstve $\hat H=H\oplus\left<\boldkey 1\right>$ zadana
struktura KTP-operatornoe0 algebry s edinitsee0 $\boldkey 1$: $l_{\vec
x}(\varphi)\boldkey 1=\exp(\vec x\cdot\vec\bold L)\varphi$, $l_{\vec
x}(\boldkey 1)=\id$.

Vmesto KTP-operatornyh algebr inogda imeet smysl rassmatrivatp1
ot\-ve\-cha\-yu\-shchie im {\cyi lokalp1nye polevye algebry}, obychnye
assotsiativnye algebry, po\-lu\-cha\-yu\-shchi\-e\-sya iz KTP-operatornyh
algebr perenormirovkoe0 potochechnogo umnozheniya operatornyh
polee0 (otmetim, chto tenzornye polya $t_{ij}^k(\vec x)$, kak
pravilo, sin\-gu\-lyar\-ny v tochke $\vec x=0$, i poe1tomu potochechnoe
proizvedenie operatornyh polee0 formalp1no beskonechno ili
neopredeleno). Protsedura perenormirovki opisana v rabote [14].
Rassmotrim dlya prostoty interesuyushchie0 nas sluchae0, kogda
$\vec x=u$ probegaet kompleksnuyu ploskostp1, a operatornye polya
meromorfny po $u$.

Rassmotrim, naryadu s operatornymi polyami $\varphi(u)$, 
parametrizuemymi e1le\-men\-ta\-mi $\varphi$ prostranstva $H$,
vyrazheniya
$$\varphi(f)=\underset u=0\to\res\left\{f(u)\varphi(u)\frac{du}u\right\}=
\lim_{u\to 0}\left\{f(u)\varphi(u)-\text{\cyr singulyarnosti}\right\}.$$
Operator $\varphi(f)$ zadaet\-sya e1lementom $\varphi$ prostranstva 
$H$ i meromorfnoe0 fun\-k\-tsi\-ee0 $f(u)$ (ili meromorfnoe0 1-formoe0
$f(u)\frac{du}u$). V silu operatornyh razlozhenie0 proizvedenie
$\varphi_{\alpha}(f)\varphi_{\beta}(g)$ dvuh operatorov
$\varphi_\alpha(f)$ i $\varphi_{\beta}(g)$ korrektno opredeleno
i dopuskaet predstavlenie
$$\varphi_{\alpha}(f)\varphi_{\beta}(g)=
\varphi_{\gamma}(h^{\gamma}_{\alpha\beta}), \text{\ \cyr gde \ }
h_{\alpha\beta}^{\gamma}(u)=\underset
v=u\to\res\left\{t_{\alpha\beta}^{\gamma}(v-u)
f(v)\frac{dv}v\right\}g(u).\tag{*}$$
Ukazannaya protsedura predstavlyaet soboe0 perenormirovku
potochechnogo proizvedeniya operatornyh polee0 v KTP-operatornoe0
algebre. Pri e1tom operatornye razlozheniya interpretiruyut\-sya
kak regulyarizatsiya potochechnogo proizvedeniya, a funktsii $f$ 
i $g$ -- kak parametry, ot kotoryh zavisit rezulp1tat perenormirovki.
Vliyanie zameny funktsionalp1nyh parametrov na rezulp1tat operatsii
pe\-re\-nor\-mi\-rov\-ki (perenormirovochnaya invariantnostp1)
opisyvaet\-sya formulami (*).

Operatory $\varphi(f)$ zamknuty otnositelp1no umnozheniya i obrazuyut
as\-so\-tsi\-a\-tiv\-nuyu algebru $\frak A(H)$. E1ta assotsiativnaya algebra
nazyvaet\-sya lokalp1noe0 polevoe algebroe0, otvechayushchee0
KTP-operatornoe0 algebre $H$. Kak pravilo, nekommutativnuyu 
lokalp1nuyu polevuyu algebru $\frak A(H)$ dlya meromorfnoe0
KTP-operatornoe0 algebry $H$ mozhno rassmatrivatp1 kak
strukturnoe kolp1tso nekotorogo nekommutativnogo mnogoobraziya
(nekommutativnogo rassloeniya nad $\Bbb C\Bbb P^1$ ili
nekommutativnogo nakrytiya $\Bbb C\Bbb P^1$) (sr.[15-17]), 
a tem samym interpretirovatp1 sovokupnostp1 operatornyh
metodov v teorii interaktivnyh dinamicheskih videosistem
realp1nogo vremeni kak {\cyi nekommutativnuyu nachertatelp1nuyu
geometriyu\/} (sr.[13]).

\subhead\cyb 2.2. Teoretiko-gruppovye i algebraicheskie
aspekty interaktivnyh dinamicheskih videosistem\endsubhead
\cyr Kak pravilo, razumno rassmatrivatp1 te konkretnye modeli
interaktivnyh dinamicheskih videosistem, kotorye obladayut toe0
ili inoe0 formoe0 invariantnosti po otnosheniyu k geometricheskim
preobrazovaniyam izobrazheniya ili vnutrennim preobrazovaniyam
tsvetovogo prostranstva. Odnoe0 iz prostee0shih form 
geometricheskoe0 invariantnosti yavlyaet\-sya invariantnostp1
otnositelp1no proektivnyh preobrazovanie0 $\Bbb C\Bbb P^1$, t.e.
gruppy $\PSLTWO$ ili ee algebry Li $\sltwo$, hotya realp1no v
silu teh ili inyh prichin, obuslovlennyh konkretnoe0
realizatsiee0, mogut rassmatrivatp1sya modeli s
invariantnostp1yu, narushennoe0 do translyatsionnyh i
masshtabnyh preobrazovanie0 (primer tomu -- obrezanie
uglovogo polya iz raboty [1]). nalichie proektivnoe0 invariantnosti
modeli oznachaet invariantnostp1 e1volyutsionnyh uravnenie0 
(formul E1e0lera, E1e0lera-Belavkina-Kolokolp1tsova i t.d.),
chto v svoyu ocheredp1 zadaet usloviya na operatory i operatornye 
polya, vhodyashchie v e1ti uravneniya. Proektivnaya invariantnostp1
operatornyh polee0 pozvolyaet v znachitelp1noe0 stepeni
spetsifitsirovatp1 ih vid (v chastnosti, analiticheski opisatp1 ih
zavisimostp1 ot $u$ i $\dot u$), chto bylo prodelano v [12]. S
drugoe0 storony, proektivnaya invariantnostp1 operatornyh polee0
vlechet proektivnuyu invariantnostp1 vseh postroennyh s ih
pomoshchp1yu algebraicheskih struktur (KTP-operatornyh algebr i
lokalp1nyh polevyh algebr). Sistematicheskie0 analiz 
sootvet\-stvuyushchih proektivno--in\-va\-ri\-ant\-nyh struktur
(operatornyh algebr kvantovoe0 proektivnoe0 teorii polya i
lokalp1nyh proektivnyh polevyh algebr) byl dan v statp1e [18].
Izlozhim kratko neobhodimye operdeleniya i rezulp1taty, sleduya
[18,19].

\definition{\cyb Opredelenie 4B {\rm [18,19]}} \cyr KTP-operatornaya
algebra $(H,t^k_{ij}(u); u\in\Bbb C)$ nazyvaet\-sya {\cyi 
KPTP-operatornoe0 algebroe0\/} ({\cyi operatornoe0 algebroe0
kvantovoe0 proektivnoe0 teorii polya\/}), esli (1) prostranstvo 
$H$ yavlyaet\-sya summoe0 modulee0 Verma $V_{\alpha}$ nad 
$\sltwo$ s e1kstremalp1nymi vektorami $v_{\alpha}$ i
e1kstremalp1nymi vesami $h_{\alpha}$, (2) $l_u(v_{\alpha})$ 
yavlyaet\-sya pervichnym polem spina $h_{\alpha}$, t.e.
$[L_k,l_u(v_\alpha)]=(-u)^k(u\partial_u+(k+1)h_{\alpha})l_u(v_{\alpha})$,
gde $L_k$ -- $\sltwo$--generatory ($[L_i,L_j]\!=\!(i\!-\!j)L_{i+j}$,
$i,j=-1,0,1$), (3) imeet mesto pravilo porozhdeniya potomkov:
$L_{-1}l_u(f)=l_u(L_{-1}f)$. KTP-operatornaya al\-geb\-ra $(H,w^k_{ij}(u);
u\in\Bbb C)$ nazyvaet\-sya {\cyi proizvodnoe0 KPTP-operatornoe0
algebroe0}, esli vypolnyayut\-sya usloviya (1) i (2) vmeste s
proizvodnym pravilom porozhdeniya potomkov: $[L_{-1},l_u(f)]=l_u(L_{-1}f)$.
\enddefinition

\cyr Kak pokazano v rabote [19] kategorii KPTP-operatornyh algebr 
i proizvodnyh KPTP-operatornyh algebr e1kvivalentny. Yavnaya
konstruktsiya e1k\-vi\-va\-lent\-nos\-ti predp2yavlena tam zhe. Kak
sledstvie, KPTP-operatornye algebry i pro\-iz\-vod\-nye KPTP-operatornye
algebry mogut rassmatrivatp1sya kak raznye zapisi odnogo i togo zhe
obp2ekta, i v konkretnoe0 situatsii mozhet ispolp1zovatp1sya
naibolee udobnaya iz nih.

Neobhodimo otmetitp1, chto v KPTP-operatornyh algebrah s 
edinitsee0 $L\!=\!\ad(L_{-1})$, a v proizvodnyh KPTP-operatornyh
algebrah $L\!=\!L_{-1}$. Primery KPTP-operatornyh algebr
rassmatrivalisp1 v [18,1,13].

Vazhnuyu rolp1 v strukturnoe0 teorii KPTP-operatornyh algebr 
igraet odna konkretnaya KPTP-operatornaya algebra --- algebra
$\Vrt(\sltwo)$ vershinnyh ope\-ra\-to\-rov dlya algebry Li $\sltwo$,
konstruktsiya kotoroe0 soderzhit\-sya v [18,19]. E1ta algebra 
realizuet\-sya v skladne modulee0 Verma nad algebroe0 Li $\sltwo$ 
i dopuskaet tochnoe predstavlenie operatornymi polyami v modeli
modulee0 Verma nad $\sltwo$. Znachenie e1toe0 algebry vyyavlyaet\-sya
sleduyushchee0 teoremoe0.

\proclaim{\cyb Teorema 1 {\cyr (sm.[18])}} \cyi Proizvolp1naya
abstraktnaya KPTP-operatornaya algebra mo\-zhet bytp1 realizovana
kak podalgebra v algebre $\Mat_n(\Vrt(\sltwo))$ matrits s
ko\-e1f\-fi\-tsi\-en\-ta\-mi v $\Vrt(\sltwo)$ pri nekotorom $n$ (konechnom,
esli KPTP-operatornaya algebra soderzhit konechnoe chislo pervichnyh
polee0, i beskonechnom v protivnom sluchae).
\endproclaim

\cyr Algebra $\Vrt(\sltwo)$ vershinnyh operatorov dlya algebry Li
$\sltwo$ obladaet ryadom dopolnitelp1nyh interesnyh svoe0stv.
tak, naprimer, pervichnye polya v e1toe0 algebre obrazuyut
beskonechnomernuyu algebru Zamolodchikova [18].

Bolee podrobno s e1toe0 algebroe0, drugimi KPTP-operatornymi
algebrami, lokalp1nymi proektivnymi polevymi algebrami (LPPA),
ih strukturnoe0 teo\-ri\-ee0, a takzhe svyazyami perechislennyh
obp2ektov s klassicheskimi al\-geb\-ra\-i\-ches\-ki\-mi strukturami mozhno
poznakomitp1sya po rabotam [18,19].

Pomimo prostranstvennyh simmetrie0 modeli interaktivnyh 
dinamicheskih videosistem mogut obladatp1 vnutrennimi 
simmetriyami, naprimer, tsvetovymi [1]. Esli e1ti simmetrii
operdeleyayut\-sya gruppoe0 $G$, to sootvet\-stvuyushchie
e1vo\-lyu\-tsi\-on\-nye uravneniya dolzhny bytp1 $G$--invariantnymi,
a algebraicheskie struktury, harakterizuyushchie e1ti
uravneniya, dopuskatp1 $G$ v kachestve gruppy simmetrie0.
Sootvet\-stvuyushchie mehanizmy podrobno issledovalisp1 v 
rabote [13] (sm.takzhe [1]). Osnovnymi al\-geb\-ra\-i\-ches\-ki\-mi
obp2ektami v e1tom sluchae budut t.n. proektivnye
$G$--gipermulp1tiplety, opredelenie kotoryh daet\-sya nizhe.

\definition{\cyb Opredelenie 4V {\rm [13]}} \cyr KPTP--operatornaya
algebra (proizvodnaya KPTP--ope\-ra\-tor\-naya algebra) $(H,t^k_{ij}(u))$
nazyvaet\-sya {\cyi proektivnym $G$--gipermulp1tipletom}, esli
gruppa $G$ dee0stvuet v nee0 avtomorfizmami, inymi slovami,
prostranstvo $H$ dopuskaet strukturu predstavleniya gruppy $G$,
operatory predstavleniya $T(g)$ kommutiruyut s dee0stviem 
$\sltwo$ i $l_u(T(g)f)=T(g)l_u(f)T(g^{-1})$.
\enddefinition

\cyr Linee0nye prostranstva e1kstremalp1nyh vektorov zadannogo
vesa obrazuyut podpredstavleniya gruppy $G$, kotorye nazyvayut\-sya
{\cyi mulp1tipletami\/} proektivnogo $G$--gipermulp1tipleta.
Proektivnye $G$--gipermulp1tiplety s $G\supseteq\SU(3)$ opisyvayut
algebraicheskuyu strukturu tsvetovogo prostranstva v
proektivno-in\-va\-ri\-ant\-nyh IAVR [1]. Konkretnye proektivnye
$G$--gipermulp1tiplety (kanonicheskie i giperkanonicheskie,
i ih chastnye sluchai -- $q_R$--konformnye teorii polya) podrobno
issledovalisp1 v rabotah [13,1,18]. Otmetim takzhe, chto mnogie
metody horosho razrabotannoe0 kvantovoe0 konformnoe0 teorii polya
neposredstvenno ili posle nekotoroe0 modifikatsii okazyvayut\-sya
primenimymi i k perechislennym konkretnym proektivnym
$G$--gipermulp1tipletam [13,18] (v chastnosti, k nim ot\-no\-syat\-sya
konstruktsiya Sugavary, polya Fubini-Venetsiano i verteksnye
konstruktsii, {\rm Virasoro master equations}, modulyarnye0 funktor
i t.d.).

\head\cyb\S 3. Organizatsiya integrirovannyh interaktivnyh
videokognitivnyh sistem realp1nogo vremeni\endhead

\cyr Dannye0 paragraf posvyashchen obshchim printsipam
organizatsii integrirovannyh interaktivnyh videokognitivnyh
sistem realp1nogo vremeni. Integrirovannostp1 ponimaet\-sya kak
sovmestnoe funktsionirovanie dvuh podsistem, odna iz kotoryh
-- iskusstvennaya (e1to, kak pravilo, intraktivnaya 
dinamicheskaya videosistema), a drugaya -- estestvennaya 
(sensornaya, v t.ch. vizualp1naya, respiratornaya, kognitivnaya
ili smeshannaya). Neobhodimo otmetitp1, chto bolp1shaya chastp1
re\-alp1\-nyh videosistem, tak ili inache nosit yavnye0 ili skrytye0
vi\-deo\-kog\-ni\-tiv\-nye0 integrirovannye0 harakter, poe1tomu predstavlyaet
interes izuchenie imen\-no integrirovannyh interaktivnyh sistem.

Odnako, prezhde chem peree0ti k izucheniyu integrirovannyh 
interaktivnyh sistem, neobhodimo datp1 analiz estestvennyh 
interaktivnyh sistem, dlya togo chtoby ponimatp1 
zakonomernosti ih funktsionirovaniya v sostave integrirovannyh
sistem. Pri e1tom, razumno rassmatrivatp1 prostee0shie sistemy
sensornogo tipa, zatem peree0ti k sootvetstvuyushchim
integrirovannym sistemam (videosensornym i videorespiratornym),
i lishp1 posle e1togo skontsentrirovatp1 vnimanie na
sobstvenno videokognitivnyh sistemah.

\subhead\cyb 3.1. Virtualizatsiya kak metod opisaniya klassa
fizicheskih interaktivnyh informatsionnyh sistem [18,20]\endsubhead
\cyr Izlozhim sleduya [18,20] osnovnye printsipy virtualizatsii
estestvennyh fizicheskih interaktivnyh sistem kak metoda ih
kinematicheskogo opisaniya.

\definition{\cyb Opredelenie 5 {\rm [20]}} \cyr {\cyi Izobrazheniem
estestvennoe0 interaktivnoe0 sistemy s pomoshchp1yu iskusstvennoe0
interaktivnoe0 videosistemy\/} budem nazyvatp1 zadanie al\-go\-rit\-ma
postroeniya dinamicheskogo izobrazheniya proizvolp1nogo
protekayushchego v es\-test\-ven\-noe0 sisteme interaktivnogo protsessa
posredstvom intentsionalp1noe0 ano\-malp1\-noe0 virtualp1noe0
realp1nosti, realizuemoe0 kompp1yuterograficheskim in\-ter\-fee0\-som
iskusstvennoe0 interaktivnoe0 videosistemy. {\cyi Virtualizatsiee0
es\-tes\-t\-ven\-noe0 interaktivnoe0 sistemy\/} budem nazyvatp1
sopostavlenie ee0 ee izob\-ra\-zhe\-niya s pomoshchp1yu iskusstvennoe0
interaktivnoe0 videosistemy po opredelennoe0 so\-vo\-kup\-nos\-ti
e1ksperimentalp1nyh dannyh o sisteme v ustanovivshemsya 
avtokolebatelp1nom rezhime. V svoyu ocheredp1 is\-hodnuyu
interaktivnuyu sistemu budem nazyvatp1 {\cyi realizatsiee0
iskusstvennoe0 interaktivnoe0 sistemy}.
\enddefinition

\cyr Inymi slovami, virtualizatsiya estestvennoe0 interaktivnoe0
sistemy poz\-vo\-lya\-et po harakteristikam nekotorogo ogranichennogo
mnozhestva av\-to\-ko\-le\-ba\-telp1\-nyh interaktivnyh protsessov
formirovatp1 dinamicheskoe izobrazhenie pro\-iz\-volp1\-no\-go (ne
obyazatelp1no avtokolebatelp1nogo) interaktivnogo protsessa
posredstvom nekotoroe0 iskusstvennoe0 interaktivnoe0 sistemy.
Vybor termina ``vir\-tu\-a\-li\-za\-tsiya'' cvyazan s tem, chto
dinamicheskoe izobrazhenie interaktivnogo pro\-tses\-sa predstavlyaet
soboe0 anomalp1nuyu virtualp1nuyu realp1nostp1 v shirokom
smys\-le e1togo ponyatiya.

Vvidu deleniya interesuyushchih nas estestvennyh interktivnyh
sistem na aktivnuyu i passivnuyu sostavlyayushchie (aktivnogo
i passivnogo agenta) harakteristiki ustanovivshegosya 
avtokolebatelp1nogo protsessa podrazdelyayut\-sya na gra\-fi\-ches\-kie
dannye o dinamicheskom sostoyanii passivnogo agenta i amplitudno-chastotnye
harakteristiki aktivnogo agenta. Virtualizatsiya estestvennoe0 
interaktivnoe0 sistemy zaklyuchaet\-sya v zadanii algoritma {\cyi
vtorichnogo sinteza\/} [20] graficheskih dannyh o dinamicheskom
sostoyanii passivnogo agenta pri lyubom interaktivnom protsesse
po zadannoe0 sovokupnosti ukazannyh dannyh pri ustanovivshihsya
avtokolebatelp1nyh protsessah iz nekotorogo mnozhestva, a takzhe
sootvet\-stvuyushchih amplitudno-chastotnyh harakteristik aktivnogo agenta
v e1tih protsessah; printsipialp1naya strukturno-algoritmicheskaya
s\-hema virtualizatsii estestvennyh interaktivnyh sistem, osnovannoe0
na ispolp1zovanii vtorichnogo sinteza izobrazhenie0 privedena v 
rabote [20].

{\cyi Kriterie0 adekvatnosti izobrazheniya i korrektnosti 
virtualizatsii {\rm [20].}} Virtualizatsiya estestvennoe0
interaktivnoe0 sistemy yavlyaet\-sya korrektnoe0, esli
po\-lu\-cha\-yu\-shche\-e\-sya v rezulp1tate nee izobrazhenie e1toe0
sistemy posredstvom nekotoroe0 iskusstvennoe0 interaktivnoe0
videosistemy adekvatno is\-hodnoe0 sisteme, t.e. pri 
odnovremennom funktsionirovanii kak iskusstvennoe0, tak i
estestvennoe0 interaktivnyh sistem s odnim i tem zhe aktivnym
agentom vosproizvedenie izobrazheniya zadannogo interaktivnogo
protsessa pozvolyaet (pri opredelennom so\-sto\-ya\-nii i reshenii
aktivnogo agenta) vyzvatp1 ego upavlyaemoe0 protekanie v
estestvennoe0 inetarktivnoe0 sisteme.

Razumeet\-sya, izobrazhenie mozhet bytp1 odnokratno postroeno 
na osnovanii e1k\-s\-pe\-ri\-men\-talp1\-nyh dannyh o funktsionirovanii
estestvennoe0 interaktivnoe0 sistemy s drugim aktivnym agentom
nezheli tot, kotorye0 uchastvuet v vosproizvedenii izobrazheniya.
Otmetim takzhe, chto pri vosproizvedenii izobrazheniya 
predusmatrivaet\-sya sovmestnoe odnovremennoe funktsionirovanie
dvuh razlichnyh interaktivnyh sistem (iskusstvennoe0 i
estestvennoe0) s odnim subp2ektom. Takim obrazom, osnovnoe
otlichie virtualizatsii ot matematicheskoe0 modeli zaklyuchaet\-sya
v tom, chto poslednyaya yavlyaet\-sya {\cyi chistym deskriptorom\/}
fizicheskogo protsessa, v to vremya kak pervaya -- {\cyi
deskriptorom--konstruktorom}.

Sformuliruem ryad feonomenologicheskih printsipov, vypolnenie
kotoryh, po-vidimomu, neobhodimo dlya korrektnosti
virtualizatsii:\newline
-- {\cyi Printsip izostrukturnoe0 ideografichnosti virtualizatsii
{\rm [20]:}} Virtualizatsiya estestvennoe0 interaktivnoe0 sistemy
dolzhna bytp1 postrena tak, chto po\-lu\-cha\-yu\-shche\-e\-sya posredstvom
nee izobrazhenie imelo tu zhe vnutrennyuyu algebraicheskuyu
strukturu, chto i is\-hodnaya sistema;\newline
-- {\cyi Printsip dinamicheskoe0 e1kvivalentnosti intentsie0
{\rm [20]:}} Virtualizatsiya es\-test\-ven\-noe0 interaktivnoe0 sistemy
dolzhna bytp1 postroena tak, chtoby parametry, harakterizuyushchie
intensivnostp1 intentsii subp2ekta na obp2ekt, dlya nee i ee
izobrazheniya sovpadali v protsesse dinamicheskogo 
subp2ekt-obp2ektnogo vza\-i\-mo\-dee0\-st\-viya. V kachestve podobnogo
parametra mozhno rassmatrivatp1 korrelyatsiyu povedencheskih
reaktsie0 subp2ekta i dinamicheskih harakteristik obp2ekta.

\subhead\cyb 3.2. Primery organizatsii integrirovannyh interaktivnyh
sistem i perspektivy razvitiya [20]\endsubhead \cyr Interaktivnye
sistemy podrazdelyayut\-sya na sle\-du\-yu\-shchie klassy:
\roster
\item"--" {\cyi Po prois\-hozhdeniyu:\/} na iskusstvennye, estestvennye
i integrirovannye.
\item"--" {\cyi Po tipu interfee0sa\/} na videosistemy, audiosistemy,
obshchie sensornye (vklyuchaya taktilp1nye), respiratornye i
kompleksnye (audiovideosistemy, videosensornye, videorespiratornye
i t.d.).
\item"--" {\cyi Po tseli funktsionirovaniay:\/} na sistemy upravleniya,
sistemy nablyudeniya, obrabotki i hraneniya informatsii,
sistemy samoregulyatsii i autotreninga, sistemy kommunikatsii.
\item"--" {\cyi Po chislu polp1zovatelee0:\/} na odno- i
mnogopolp1zovatelp1skie.
\endroster

K naibolee interesnym otnosyat\-sya sredi mnogih drugih 
sleduyushchie vidy interaktivnyh sistem: (1) iskusstvennye
videosistemy ``avtomaticheskoe0 aranzhirovki vospriyatiya'',
v tom chisle mnogopolp1zovatelp1skie, ispolp1zuyushchie
vto\-rich\-nye0 sintez izobrazhenie0 (VSI) [20], (2) vklyuchayushchie
predydushchie0 vid v kachestve podsistemy integrirovannye
videosensornye sistemy interaktivnogo zreniya, (3) integrirovannye
videorespiratornye sistemy samoregulyatsii, autotreninga i
kognitivnoe0 stimulyatsii, (4) integrirovannye 
mno\-go\-polp1\-zo\-va\-telp1\-s\-kie videokognitivnye sistemy uskorennoe0
neverbalp1noe0 kommunikatsii.

Perspektivnym v oblasti sistem VSI [20], igrayushchih klyuchevuyu
rolp1 pri proektirovanii mnogih integrirovannyh sistem,
yavlyaet\-sya, po-vidimomu, izuchenie intellektualp1nyh sistem
VSI s dinamicheskoe0 interaktivnoe0 nastroe0koe0 (a takzhe
razrabotka portativnyh videosistem sinhronnogo VSI i kompleksnyh
mnogopolp1zovatelp1skih sistem). Sistemy VSI s dinamicheskoe0
interaktivnoe0 nastroe0koe0 mogut rassmatrivatp1sya kak sistemy
``avtomaticheskoe0 aranzhirovki vospriyatiya'' v rakurse
deyatelp1nosti, svyazannoe0 s kompp1yuternymi sistemami
interaktivnogo ``avtomaticheskogo pisp1ma'', a takzhe issledovanie0
sinteticheskoe0 pertseptsii. Imeet smysl i izuchenie oktonionnogo
anomalp1nogo {\rm 3D} stereosinteza [1], kiralp1noe0
dissimmetrii zritelp1nogo analizatora v interaktivnyh protsessah
i ee vliyaniya na stereosintez. Predstavlyaet\-sya vazhnym
izuchenie sovmestnogo funktsionirovaniya estsevennyh interaktivnyh
sistem i ih vir\-tu\-a\-li\-za\-tsie0 pri razrabotke integrirovannyh
interaktivnyh sistem ( realizuyushchih {\cyi ``integrirovannuyu
realp1nostp1''\/} kak alp1ternativu sistemam ``virtualp1noe0
re\-alp1\-nos\-ti''). K integrirovannym sistemam, vklyuchayushchim
v sebya iskusstvennye interaktivnye videopodsistemy 
avtomaticheskoe0 aranzhirovki vospriyatiya, ot\-no\-syat\-sya
integrirovannye videosensornye sistemy interaktivnogo zreniya 
s es\-test\-ven\-noe0 sensornoe0 podsistemoe0 i integrirovannye
videorespiratornye interaktivnye sistemy (t.e. interaktivnye 
videosistemy s respiratornoe0 mo\-du\-lya\-tsi\-ee0 ili, v bolee obshchem
sluchae, transformatsiee0 vizualp1nyh interaktivnyh protsessov)
psihofiziologicheskogo autotreninga (kak 
passivno-re\-lak\-sa\-tsi\-on\-no\-go, tak i aktivno dinamicheskogo).

\subhead\cyb 3.3. Organizatsiya integrirovannyh videokognitivnyh
sistem realp1nogo vremeni dlya uskorennoe0 neverbalp1noe0
kognitivnoe0 kommunikatsii\endsubhead \cyr Ne\-ko\-to\-rye kognitivnye
aspekty interaktivnoe0 kompp1yuternoe0 grafiki v tselom
ras\-smat\-ri\-va\-lisp1 v monografii [21]. V dannoe0 rabote my
interesuemsya kog\-ni\-tiv\-ny\-mi aspektami kommunikatsionnyh
interaktivnyh videosistem realp1nogo vremeni (a takzhe i ih
integrirovannyh videosensornyh analogov).

Kak otmechalosp1 ranee, neposredstvennoe ispolp1zovanie
IAVR-telestezii, po-vidimomu, ne pozvolyaet osushchestvlyatp1
peredachu kognitivnoe0 informatsii. S drugoe0 storony,
integrirorvannye videokognitivnye kommunikatsionnye
interaktivnye sistemy mogut ``nasledovatp1'' poleznye svoe0stva
svoih videopodsistem, realizuyushchih IAVR-telesteziyu, no 
pri e1tom peredavatp1 i slozhnuyu kognitivnuyu informatsiyu.
Tak, podhod k razrabotke podobnyh ``proizvodnyh'' integrirovannyh
sistem uskorennoe0 neverbalp1noe0 kognitivnoe0 kommunikatsii,
baziruyushchihsya na primenenii IAVR-telestezii (sm.vyshe) v
kachestve apparata predvaritelp1nogo raspoznavaniya i
interaktivnoe0 {\cyi samonastroe0ki\/} kom\-mu\-ni\-ka\-tsi\-on\-nyh
``klyuchee0'' subp2ektov, a takzhe ih dinamicheskoe0
sinhronizatsii, prizvan obespechitp1 pokanalp1nuyu interaktivnuyu
opredelyaemostp1 ``klyuchee0'' svyazi i, takim obrazom,
printsipialp1nuyu neinterpretiruemostp1 (nedeshifruemostp1)
soobshcheniya izvne. Realizuemostp1 podobnoe0 s\-hemy opiraet\-sya
na ryad ob\-s\-to\-ya\-telp1stv: (1) nalichie interpretatsionnyh figur,
nablyudaemyh tolp1ko v mnogopolp1zovatelp1skom rezhime, 
naryadu s drugimi osobennostyami mno\-go\-polp1\-zo\-va\-telp1s\-ko\-go
rezhima v interpretatsionnoe0 geometrii, obsuzhdavshimisya 
vyshe (na\-pri\-mer, polisemantichnostp1yu i kvantovologicheskim
harakterom kom\-mu\-ni\-ka\-tsii, pozvolyayushchim primenyatp1 nekotorye
idei i metody kvantovoe0 kriptografii [22-25]), (2) sposobnostp1
interpretatsionnyh figur sluzhitp1 ukazatelyami na
vi\-deo\-kog\-ni\-tiv\-nye obp2ekty otlichnoe0 ot nih prirody, dinamicheski
re\-kon\-st\-ru\-i\-ru\-e\-mye v realp1nom vremeni polp1zovatelyami v
protsesse kommunikatsii. Ukazannye obp2ekty budem nazyvatp1
{\cyi dinamicheski rekonstruiruemymi obp2ektami
e1k\-s\-pe\-ri\-men\-talp1\-noe0 matematiki} ili kratko {\cyi ``droe1mami''}.
Takim obrazom, potok vi\-zu\-alp1\-noe0 informatsii mezhdu polp1zovatelyami,
analizirovavshie0sya v \S1.4, or\-ga\-ni\-zu\-et bolee slozhnye0 potok
videokognitivnoe0 informatsii, voznikayushchie0 mezh\-du nimi
(otmetim, chto hotya peredavaemaya informatsiya kognitivna,
protsess ee peredachi nosit interaktivno upravlyaemye0 i, otchasti,
bessoznatelp1nye0 harakter, prichem dvustironnie0 potok 
interaktivnoe0 videoinformatsii mozhet ras\-smat\-ri\-vatp1\-sya kak
psihosemanticheskie0 kontekst kognitivnogo informatsionnogo 
obmena). Pri e1tom is\-hodnaya interpretatsionnaya geometriya
(tochnee, interaktivnaya videosistema -- ee ``nositelp1'') sluzhit
virtualizatsiee0 (deskriptorom-konstruktorom) realizuemoe0
kognitivnoe0 interaktivnoe0 sistemy. K obsuzhdeniyu 
matematicheskih aspektov temy, a imenno, droe1mam i ih
dinamicheskoe0 rekonstruktsii my peree0dem v sleduyushchem
paragrafe.

\head\cyb\S 4. Droe1my i ih dinamicheskaya rekonstruktsiya\endhead

\cyr Dannye0 paragraf posvyashchen droe1mam i ih dinamicheskoe0
rekonstruktsii. Ho\-tya, v tselom, ukazannye0 protsess, kak
vprochem i obshchaya priroda droe1mov, ne vpolne yasny dazhe na
kontseptualp1nom urovne (hotya, po-vidimomu, droe1my kak-to
svyazany s t.n. dinamicheskimi simulyakrami ponyatie0 [20:Prim.5]),
otdelp1nye primery mogut bytp1 razobrany dostatochno formalp1no,
tem samym proyasnyaya nekotorye obshchie mehanizmy. Primer, 
kotorye0 podrobno razbiraet\-sya nizhe, svyazan s
beskonechnomernoe0 geometriee0, i mozhet bytp1 oharakterizovan
kak otvet na vopros ``mozhno li nablyudatp1 beskonechnomernye
obp2ekty?'', inymi slovami, vozmozhna li i kak {\cyi
nachertatelp1naya beskonechnomernaya geometriya}. Kak my
uvidim ispolp1zovanie interaktivnyh videokognitivnyh sistem,
droe1mov i ih di\-na\-mi\-ches\-koe0 rekonstruktsii v opredelennom
smysle reshaet e1tot vopros polozhitelp1no v printsipe, chto 
{\cyi pozvolyaet ispolp1zovatp1 metody e1ksperimentalp1noe0
(kom\-pp1yu\-ter\-noe0) matematiki pri izuchenii beskonechnomernyh
geometricheskih obp2ektov}. Otmetim, chto nekotorye aspekty
beskonechnomernoe0 nachertatelp1noe0 geometrii teo\-re\-ti\-ches\-ki
obsuzhdalisp1 v rabote [26].

Rassmotreniyu samogo primera (pp.4.3.,4.4.) tselesoobrazno
predposlatp1 obsuzhdenie dvuh tehnicheskih voprosov (pp.4.1.,4.2.).

\subhead\cyb 4.1. Organizatsiya kiberprostranstva [7,2:Prilozh.A]\endsubhead
\cyr Pustp1 dlya prostoty dinamika izobrazhenie0 zadaet\-sya
formulami E1e0lera, kotorye spareny s nekotoro0 yavnoe0
zavisimostp1yu uglovyh polee0 ot vremeni, tak chto v tselom ih
sovokupnostp1 proektivno-invariantna. Kiberprostranstvo
sostoit iz pro\-st\-ran\-st\-va izobrazhenie0 $V_I$ s fundamentalp1noe0
dlinoe0 (shagom reshetki) $\Delta_I$ i pro\-st\-ran\-st\-va nablyudeniya
$V_O$ s fundamentalp1noe0 dlinoe0 $\Delta_O$; v pro\-st\-ran\-st\-ve $V_I$
formiruyut\-sya izobrazheniya, v to vremya kak pro\-st\-ran\-st\-vo $V_O$
ispolp1zuet\-sya dlya dannyh o dvizhenii glaz; estestvenno
potrebovatp1 $\Delta_I\gg\Delta_O$. Formuly E1e0lera zapisyvayut\-sya
v vide $\dot\Phi_t=A_{t,\xi}(u,\dot u)\Phi_t$, budem schitatp1, chto
komponenty raz\-lo\-zhe\-niya uglovogo polya $A_{t,\xi}(u,\dot u)$ po
$\dot u$ porozhdayut KPTP-operatornuyu algebru $q_R$--konformnoe0
teorii polya [18], t.e. yavlyayut\-sya $\sltwo$--pervichnymi
polyami v mo\-du\-le Verma $V_h$ nad algebroe0 Li $\sltwo$. Uglovoe
pole $A_{t,\xi}(u,\dot u)$ mozhet bytp1 pri\-bli\-zhen\-no predstavleno
v vide $M_1(t,\xi)\dot uV_1(u)+M_2(t,\xi)\dot u^2V_2(u)+\dots
+M_n(t,\xi)\dot u^nV_n(u)$, gde velichiny $M_i(t,\xi)$ realizuyut
yavnuyu zavisimostp1 uglovogo polya ot vremeni i
neogeometricheskih dinamicheskih parametrov tipa biopotentsialov
E1E1G ili dannyh respiratornogo ritma, a $V_i(u)$ yavlyayut\-sya
$\sltwo$--pervichnymi polyami spina $i$ v module Verma $V_h$ nad
algebroe0 Li $\sltwo$. Yavnye formuly dlya e1tih polee0 privedeny,
naprimer, v [16], takzhe kak ih diskretnye (reshetochnye) analogi,
kotorye i ispolp1zuyut\-sya na praktike (pri e1tom $n$ ravno 2 ili
3).

\subhead\cyb 4.2. Obrezanie uglovogo polya [1]\endsubhead \cyr Iz
prakticheskih soobrazhenie0 imeet smysl rassmatrivatp1 obrezanie
$A^{\cut}(u,\dot u)$ uglovogo polya $A(u,\dot u)$ v regulyarnoe0 
chasti, ne soderzhashchee stepenee0 $u$ s pokazatelem bolp1shim
zadannogo $N$. Predpolagaet\-sya, chto $A^{\cut}(u,\dot u)$ 
ostaet\-sya translyatsionno i masshtabno invariantnym, a
di\-la\-ta\-tsi\-on\-naya invariantnostp1 narushaet\-sya. Naprimer,
komponenty obrezannogo $q_R$--affinnogo toka $J^{\cut}(u)$
($\sltwo$--pervichnogo polya spina 1 v module Verma $V_h$ nad
algebroe0 Li $\sltwo$) zadayut\-sya operatorami
$J_k^{\cut}=J_k=\partial_z^k$, $J_{-k}^{\cut}=z^k\Delta_+^kP(z\partial_z)$
($\Delta_+f(x)=f(x+1)-f(x)$),
gde $P(z\partial_z)$ -- mnogochlen stepeni $N$, naprimer, takoe0, 
chto $P(z\partial_z)z^i=\frac1{2h+i}z^i$, $i\le N$. Modulp1 Verma
$V_h$ realizovan v prostranstve mnogochlenov ot odnoe0
peremennoe0 $z$, $\sltwo$--generatory imeyut vid 
$L_1=(z\partial_z+1h)\partial_z$, $L_0=z\partial_z+h$, $L_{-1}=z$.
Mnogochlen $P(z\partial_z)$ odnoznachno opredelyaet operator 
$L_1^{\cut}$ takoe0, chto $[L_1^{\cut},J_{-1}^{\cut}]=1$, 
$[L_1^{\cut},L_0]=L_1^{\cut}$; $L_1^{\cut}$ yavlyaet\-sya
obrezannym operatorom dilatatsii: $L_1^{\cut}=zP^{-1}(z\partial_z)$.
Operatory $L_1^{\cut}$, $L_0$, $L_{-1}$ porozhdayut t.n. nelinee0nuyu
$\operatorname{sl}_2$ [27] s sootnosheniyami
$$[L_0,L_{-1}]=L_{-1},\quad [L_1^{\cut},L_0]=L_1^{\cut},\quad
[L_1^{\cut},L_{-1}]=h(L_0),$$
gde $h(x)=\frac1{P(x+1)}-\frac1{P(x)}$. Ukazannaya nelinee0naya
$\operatorname{sl}_2$ opisyvaet narushenie proektivnoe0
invariantnosti pri protsedure obrezaniya.

Protsedura obrezaniya pri $N=1$ i $A(u,\dot u)=J(u)\dot u$ v
chastnyh sluchayah rea\-li\-zu\-et interaktivnuyu videosistemu s
chastichnym uvlecheniem i maskirovaniem, t.e.
$$\Phi(x)=\Phi_u(x)=f(|x-u|)\Phi_0(x-\gamma u),$$
gde $\gamma$ -- koe1ffitsient uvlecheniya, $f$ -- funktsiya
maskirovaniya, $u=u(t)$ -- polozhenie tochki vzora. Videodannye
mogut bytp1 mnogosloe0nymi s koe1ffitsientom uvlecheniya i
funktsiee0 maskirovaniya svoimi dlya kazhdogo sloya.

Perehod k diskretnoe0 (reshetochnoe0) versii v sluchae obrezaniya
ne vyzyvaet problem.

\subhead\cyb 4.3. Beskonechnomernye dinamicheskie simmetrii 
interaktivno upravlyaemyh videosistem\endsubhead \cyr
Beskonechnomernye dinamicheskie simmetrii interaktivno 
up\-rav\-lya\-emyh videosistem, e1volyutsiya kotoryh opisyvaet\-sya
formulami E1e0lera ili E1e0lera-Belavkina-Kolokolp1tsova 
(vozmozhno, sparennymi s uravneniyami E1e0\-le\-ra-Arnolp1da) s
operatornymi polyami, porozhdayushchimi KPTP--operatornuyu 
al\-geb\-ru $q_R$--konformnoe0 teorii polya, mogut bytp1 postroeny
odnim iz sle\-du\-yu\-shchih treh e1kvivalentnyh sposobov.

{\cyi Sposob 1\/} [18]. Generatory beskonechnomernyh
dinamicheskih simmetrie0 sutp1 $\sltwo$--tenzornye operatory
v modulyah Verma $V_h$ nad algebroe0 Li $\sltwo$, 
preobrazuyushchiesya pod dee0stviem $\sltwo$ kak golomorfnye
$n$--differentsialy v edi\-nich\-nom kompleksnom diske $D_+$
($n\in\Bbb Z_-$), t.e. $m$--polivektornye polya ($m\in\Bbb Z_+$). 
Takim obrazom, proizvodyashchie funktsii dlya generatorov
beskonechnomernyh dinamicheskih simmetrie0 sutp1 $\sltwo$--pervichnye
operatornye polya v $V_h$ spina $m$.

{\cyi Sposob 2}. Dee0stvie algebry Li $\sltwo$ v module Verma $V_h$
prodolzhaet\-sya do predstavleniya $T$ algebry Li $W_1$ formalp1nyh
vektornyh polee0 na pryamoe0 [28]. A imenno, esli modulp1 Verma $V_h$ 
realizovan v prostranstve mnogochlenov odnoe0 kompleksnoe0 
peremennoe0 $z$, a generatory $\sltwo$ imeyut vid $L_{-1}=z$, 
$L_0=z\partial_z+h$, $L_1=z\partial^2_z+2h\partial_z$, to ostalp1nye
generatory algebry Li $W_1$ zadayut\-sya operatorami
$L_k=z\partial_z^{k+1}+(k+1)h\partial_z^k$ ($k\ge 2$). Esli modulp1
Verma $V_h$ unitarizuem, to generatory beskonechnomernyh
di\-na\-mi\-ches\-kih simmetrie0, otvechayushchih $\sltwo$--pervichnomu
polyu spina 2, predstavlyayut\-sya v vide $T(X)$ ili $T^*(X)$, gde 
$X\in W_1$. V neunitarizuemom sluchae sleduet vospolp1zovatp1sya
analiticheskim prodolzheniem po parametru $h$. Chtoby poluchitp1
generatory beskonechnomernyh di\-na\-mi\-ches\-kih simmetrie0,
otvechayushchih $\sltwo$--pervichnomu polyu spina 1, neobhodimo
prodolzhitp1 dee0stvie algebry Li $W_1$ v module Verma $V_h$ do
predstavleniya $\tilde T$ polupryamoe0 summy e1toe0 algebry i
abelevoe0 algebry Li $\Bbb C[z]$ v $V_h$. E1to predstavlenie pri
ogranichenii na $\Bbb C[z]$ yavlyaet\-sya predstavleniem 
ukazannoe0 algebry ne tolp1ko kak abelevoe0 algebry Li, no i 
kak kommutativnoe0 assotsiativnoe0 algebry, obrazuyushchee0 $z$
otvechaet operator $\partial_z$. Generatory beskonechnomernyh
dinamicheskih simmetrie0, otvechayushchih $\sltwo$--pervichnomu 
polyu spina 1, imeyut vid $\tilde T(X)$ ili $\tilde T^*(X)$, 
gde $X\in\Bbb C[z]$. Pri pomoshchi analogichnoe0, no neskolp1ko
bolee slozhnoe0 konstruktsii mozhno poluchitp1 i ostalp1nye
beskonechnomernye di\-na\-mi\-ches\-kie simmetrii.

{\cyi Sposob 3}. Prodolzhim predstavlenie algebry Li $\sltwo$ 
v module Verma $V_h$ do predstavleniya $T$ algebry Li 
$\Vect^{\Bbb C}(S^1)$ gladkih $\Bbb C$--znachnyh vektornyh
polee0 na okruzhnosti $S^1$ (bolee tochno, $\Bbb Z$--graduirovannoe0
algebry Vitta polinomialp1nyh vektornyh polee0) v module
$V(h)$ funktsionalp1noe0 razmernosti 1. Oboznachim $P$
estestvennye0 $\sltwo$--invariantnye0 proektor prostranstva 
$\End(V(h))$ na prostranstvo $\End(V_h)$. Beskonechnomernye
dinamicheskie simmetrii, otvechayushchie $\sltwo$--pervichnomu
polyu spina 2, imeyut vid $P(T(X))$ ($X\in\Vect^{\Bbb C}(S^1)$). 
Chtoby poluchitp1 vse ostalp1nye beskonechnomernye di\-na\-mi\-ches\-kie
simmetrii neobhodimo rassmotretp1 algebru Li
$\DOP^{\Bbb C}_{[\cdot,\cdot]}(S^1)$ differentsialp1nyh 
operatorov vmesto algebry Li $\Vect^{\Bbb C}(S^1)$ vektornyh polee0.

Algebraicheskaya struktura opisannyh beskonechnomernyh 
dinamicheskih simmetrie0 byla raskryta v rabotah [29,30].
Dadim okonchatelp1nuyu formulirovku rezulp1tata.

\proclaim{\cyb Teorema 2A} \cyi Beskonechnomernye dinamicheskie
simmetrii, otvechayushchie $\sltwo$-per\-vich\-no\-mu polyu spina 2,
obrazuyut $\HS$-proektivnoe predstavlenie algebry Li
$\Vect^{\Bbb C}(S^1)$ v unitarizuemom module Verma $V_h$ (t.e.
predstavlenie po modulyu operatorov Gilp1berta-Shmidta), a takzhe
asimptoticheskoe predstavlenie ukazannoe0 al\-geb\-ry ``$mod\ O(\hbar)$''
(v smysle [31]; $\hbar\!=\!h\!-\!\frac12$). Vsya sovokupnostp1
beskonechnomernyh di\-na\-mi\-ches\-kih simmetrie0 obrazuet
$\HS$-proektivnoe i asimptoticheskoe ``$mod\ O(\hbar)$''
predstavleniya algebry $\DOP^{\Bbb C}_{[\cdot,\cdot]}(S^1)$.
\endproclaim

\cyr Utverzhdenie teoremy legko vyvodit\-sya iz tretp1ego 
sposoba opredeleniya bes\-ko\-nech\-no\-mernyh dinamicheskih simmetrie0.

Chastp1 iz beskonechnomernyh dinamicheskih simmetrie0 mozhet bytp1
``globalizovana'' [30]. Sformuliruem okonchatelp1nye0 rezulp1tat.

\proclaim{\cyb Teorema 2B} \cyi Beskonechnomernye dinamicheskie
simmetrii, otvechayushchie $\sltwo$--pervichnym polyam spinov 1
i 2, e1ksponentsiruyut\-sya do proektivnogo 
$\HS$--psev\-do\-pred\-s\-tav\-le\-niya [30] polupryamogo proizvedeniya
gruppy $\Diff_+(S^1)$ diffeomorfizmov okruzhnosti i gruppy petelp1
$\Map(S^1,U(1))$, yavlyayushchegosya asimptoticheskim 
predstavleniem ``$mod\ O(\hbar)$''.
\endproclaim

\cyr Teorema 2 dopuskaet analog dlya beskonechnomernyh 
dinamicheskih simmetrie0 dlya dinamik v proizvolp1nyh 
kanonicheskih $G$--gipermulp1tipletah [13,1]. Dlya e1togo
nado zamenitp1 abelevu gruppu $U(1)$ na gruppu $G$.

Rassmotrim priblizhenie uglovogo polya s $n=2$ (sm.4.1.). V e1tom
sluchae ug\-lo\-voe pole predstavlyaet\-sya v vide generatorov
beskonechnomernyh di\-na\-mi\-ches\-kih simmetrie0 s koe1ffitsientami,
zavisyashchimi ot parametrov upravleniya. Kak sledstvie, dinamika
integriruet\-sya po modulyu operatorov Gilp1berta-Shmid\-ta ili
asimptoticheski ``$mod\ O(\hbar)$'' i zadaet\-sya interaktivno
upravlyaemym gruppovym e1lementom polupryamogo proizvedeniya
gruppy diffeomorfizmov okruzhnosti i gruppy petelp1.

\subhead\cyb 4.4. Beskonechnomernye droe1my i ih dinamicheskaya
rekonstruktsiya\endsubhead \cyr Iz re\-zulp1\-ta\-tov predydushchego
punkta sleduet, chto interpretatsionnye figury v mo\-de\-lyah,
zadavaemyh kanonicheskimi proektivnymi $G$--gipermulp1tipletami,
mogut sluzhitp1 ukazatelyami na beskonechnomernye droe1my, 
realizuemye pri pomoshchi geometricheskih obp2ektov, svyazannyh
s gruppami diffeomorfizmov okruzhnosti i grupp petelp1
(sm.[26,32-35,17] i ssylki v nih). Protsess dinamicheskoe0
rekonstruktsii zaklyuchaet\-sya v vosstanovlenii beskonechnomernogo
obp2ekta po interaktivnomu protsessu v dinamicheskoe0 videosisteme.
Sam beskonechnomernye0 obp2ekt mozhet bytp1 kak interpretatsionnym,
tak i staticheskim (kompilyatsionnym). Na\-pri\-mer, interpretatsionnye
figury mogut sluzhitp1 ukazatelyami na staticheskoe ``izobrazhenie''
na ``beskonechnomernom e1krane'' -- prostranstve universalp1noe0
deformatsii kompleksnogo diska [33,34]. V kachestve droe1mov 
mogut vystupatp1 i mnogochislennye (podcherknem, podchas
vesp1ma e1kzoticheskie) beskonechnomernye obp2ekty geometricheskoe0
teorii vtorichnokvantovannyh strun [35], chto otmechalosp1 v
tretp1ee0 chasti raboty [35]; takim obrazom, imeet smysl
govoritp1 i o {\cyi nachertatelp1noe0 strunnoe0 geometrii} v
kontekste obshchego matematicheskogo formalizma teorii strun 
(sm.napr.[36]).

Otmetim, chto perehod k reshetochnoe0 versii, po-vidimomu,
dolzhen privoditp1 k kvantovym analogam beskonechnomernyh 
grupp i algebr Li (sr.[37]), odnako, geometricheskie
posledstviya protsedury obrezaniya (sm.vyshe p.4.2.)
neizvestny.

Vazhnostp1 yazyka beskonechnomernoe0 geometrii otmechalasp1 v
[26]; dannaya ra\-bo\-ta mozhet rassmatrivatp1sya kak razvitie
sformulirovannogo v [26] tezisa o estestvennosti yazyka
beskonechnomernoe0 geometrii. Otmetim, chto droe1my (v
chastnosti, beskonechnomernye) i ih dinamicheskaya
rekonstruktsiya pomimo opisannoe0 prikladnoe0 problemy 
organizatsii uskorennyh neverbalp1nyh kognitivnyh 
kom\-pp1yu\-ter\-nyh i telekommunikatsie0 buduchi vesp1ma vazhnymi
dlya e1ksperimentalp1noe0 matematiki i, kak sledstvie, dlya 
vsego kompleksa matematicheskih (v tom chisle teoreticheskih)
nauk, predstavlyayut interes i dlya teoreticheskoe0
matematicheskoe0 psihologii, strukturnoe0 lingvistiki i
lingvisticheskoe0 psihologii v kontekste kak stavshih aktualp1nymi
sravnitelp1no nedavno issledovanie0 neverbalp1nyh kognitivnyh
kommunikatsie0 v razlichnyh vneshnih usloviyah (v tom chisle
pri nalichii stimulyatsie0), tak i v kontekste takih bolee
traditsionnyh tem, kak izuchenie vozniknoveniya i razvitiya
verbalp1nyh kommunikatsie0, rannih e1tapov formirovaniya rechi
i protsessov obucheniya. V e1toe0 svyazi predstavlyaet osobye0
interes dinamicheskaya rekonstruktsiya droe1mov v
mnogopolp1zovatelp1skom rezhime (sr.p.1.4.), naprimer, v
interaktivnyh videokognitivnyh igrah (igrah s interaktivnym
upravleniem, sm.p.1.1.).

\Refs\nofrills{\cyb Spisok literatury}
\roster
\item" [1]" {\cyre Yurp1ev D.V., Oktoniony i binokulyarnoe ``podvizhnoe
videnie'' /\!/ FPM. 1998, v pechati} [Draft English e-version: hep-th/9401047
(1994)].
\item" [2]" {\cyre Yurp1ev D.V., {\rm Watch-dog} e1ffekty
Belavkina-Kolokolp1tsova v interaktivno upravlyaemyh stohasticheskih
dinamicheskih videosistemah /\!/ TMF. 1996. T.106, vyp.2. S.333-352}.
\item" [3]" Kalawsky R.S., The science of virtual reality and virtual
environments. Addison-Wesley, 1993.
\item" [4]" Virtual reality: applications and explorations. Ed.A.Wexelblat.
Acad.Publ., Boston, 1993.
\item" [5]" Burdea G., Coiffet Ph., Virtual reality technology. J.Wiley \&\
Sons, 1994.
\item" [6]" {\cyre Forman N., Vilp1son P., Ispolp1zovanie virtualp1noe0
realp1nosti v psihologicheskih issledovaniyah /\!/ Psihol.zhurn. 1996.
T.17. vyp.2. S.64-79.}
\item" [7]" Juriev D., Visualizing 2D quantum field theory: geometry and
infomatics of mobilevision: Report RCMPI-96/02 (1996) [Draft e-version:
hep-th/9401067+hep-th/9404137 (1994)].
\item" [8]" Saaty T.L., Speculating on the future of Mathematics /\!/
Appl.Math.Lett. 1988. V.1. P.79-82.
\item" [9]" Beltrametti E.G., Cassinelli G., The logic of quantum mechanics.
Encycl.Math.Appl.15, Ad\-di\-son-Wesley Publ., London, 1981.
\item"[10]" {\cyre Oue1n G., Teoriya igr. M., 1971.}
\item"[11]" {\cyre Vorobp1ev N.N., Teoriya igr. L., 1985}.
\item"[12]" {\cyre Yurp1ev D.V., Kvantovaya proektivnaya teoriya polya:
kvantovo-polevye analogi formul E1e0lera /\!/ TMF. 1992. T.92, vyp.1.
S.172-176.}
\item"[13]" {\cyre Yurp1ev D.V., Kvantovaya proektivnaya teoriya polya:
kvantovo-polevye analogi uravnenie0 E1e0lera-Arnolp1da v proektivnyh
$G$-gipermulp1tipletah /\!/ TMF. 1994. T.98, vyp.2. S.220-240.}
\item"[14]" {\cyre Yurp1ev D.V., KPTP-operatornye algebry i
kommutativnoe vneshnee dif\-fe\-ren\-tsi\-alp1\-noe is\-chis\-le\-nie
/\!/ TMF. 1992. T.93, vyp.1. S.32-38.}
\item"[15]" Witten E., Non-commutative geometry and string field theory /\!/
Nucl.Phys.B 1986. V.268. P.253-291.
\item"[16]" Witten E., Quantum field theory, grassmannians and algebraic
curves /\!/ Commun.Math.Phys, 1988. V.113. P.529-600.
\item"[17]" {\cyre Yurp1ev D.V., Kvantovaya konformnaya teoriya polya kak
beskonechnomernaya ne\-kom\-mu\-ta\-tiv\-naya geometriya /\!/ UMN. 1991.
T.46, vyp.4. S.115-138.}
\item"[18]" {\cyre Yurp1ev D.V., Kompleksnaya proektivnaya geometriya i
kvantovaya proektivnaya teoriya polya /\!/ TMF. 1994. T.101, vyp.3.
S.331-348.}
\item"[19]" {\cyre Bychkov S.A., Yurp1ev D.V., Tri algebraicheskie struktury
kvantovoe0 proektivnoe0 ($\sltwo$--invariantnoe0) teorii polya /\!/ TMF. 1993.
T.97, vyp.3. S.336-347.}
\item"[20]" {\cyre Yurp1ev D.V., K opisaniyu klassa fizicheskih interaktivnyh
informatsionnyh sistem}: Report RCMPI-96/05 (1996) [e-version:
mp\_arc/96-459 (1996)].
\item"[21]" {\cyre Zenkin A.A., Kognitivnaya kompp1yuternaya grafika.
M., Nauka, 1991.}
\item"[22]" Wiesner S., Conjugate coding /\!/ SIGACT News. 1983. V.15, no.1. 
P.78-88.
\item"[23]" Wiedemann D., Quantum cryptography /\!/ SIGACT News. 1989.
V.18, no.2. P.28-30.
\item"[24]" Bennett C.H., Brassard G., The dawn of a new era for quantum
cryptography: the experimental prototype is working /\!/ SIGACT News. 1989.
V.20, no.2. P.78-82.
\item"[25]" Bennett C.H., Brassard G., Cr\'epeau C., Jozsa R., Peres A.,
Wootwers W.K., Teleporting and unknown quantum state via dual classical and
EPR channels /\!/ Phys.Rev.Lett. 1993. V.70. P.1895-1899.
\item"[26]" Juriev D., The vocabulary of geometry and harmonic analysis on the
infinite--dimensional manifold $\Diff_+(S^1)/S^1$ /\!/ Adv.Soviet Math. 1991.
V.2. P.233-247.
\item"[27]" Ro\v cek M., Representation theory of the nonlinear $\SU(2)$
algebra /\!/ Phys.Lett.B. 1991. V.255. P.554-557.
\item"[28]" {\cyre Fuks D.B., Kogomologii beskonechnomernyh algebr Li.
M., Nauka, 1983.}
\item"[29]" Juriev D., Topics in hidden symmetries. V. E-print:
funct-an/9611003.
\item"[30]" Juriev D., On the infinite-dimensional hidden symmetries. I-III.
E-prints: funct-an/9612004, funct-an/9701009, funct-an/9702002.
\item"[31]" {\cyre Karasev M.V., Maslov V.P., Nelinee0nye skobki Puassona.
Geometriya i kvantovanie. M., Nauka, 1991.}
\item"[32]" {\cyre Yurp1ev D.V., Neevklidova geometriya zerkal i
predkvantovanie na odnorodnom ke1\-le\-ro\-vom mnogoobrazii
$M=\Diff_+(S^1)/\operatorname{Rot}(S^1)$ /\!/ UMN. 1988. T.43. vyp.2.
S.187-188.}
\item"[33]" {\cyre Yurp1ev D.V., Modelp1 modulee0 Verma nad algebroe0 Virasoro
/\!/ Algebra i anal. 1990. T.2. vyp.2. S.209-226.}
\item"[34]" Juriev D., Infinite--dimensional geometry of the universal
deformation of the complex disk /\!/ Russian J.Math.Phys. 1994. V.2.
P.111-121.
\item"[35]" Juriev D., Infinite dimensional geometry and quantum field theory
of strings. I-III /\!/ Alg.Groups Geom. 1994. V.11. P.145-179 [e-version:
hep-th/9403068 (1994)]; Russian J.Math.Phys. 1996. V.4. P.287-314;
J.Geom.Phys. 1995. V.16. P.275-300.
\item"[36]" {\cyre Grin M., Shvarts Dzh., Vitten E1., Teoriya superstrun. M.,
Mir, 1990.}
\item"[37]" Reshetikhin N.Yu., Semenov-Tian-Shansky M.A., Central extensions
of quantum current groups /\!/ Lett.Math.Phys. 1990. V.19. P.133-142.
\endroster
\endRefs
\enddocument